\documentclass[aps,prb,twoside,twocolumn,notitlepage,nofootinbib,floatfix,floats,amssymb,amsmath]{revtex4-1}

\usepackage{color}
\usepackage{amsmath}
\usepackage{pifont}  
\usepackage{graphicx}
\usepackage{dcolumn} 
\usepackage{bm}      
\usepackage{amsfonts}
\usepackage{amssymb} 

\graphicspath{{./}{./FIG/}}

\begin{document}

\newcommand{\ba}{{\bf a}}
\newcommand{\bbb}{{\bf b}}
\newcommand{\BB}{{\bf b}}
\newcommand{\bd}{{\bf d}}
\newcommand{\br}{{\bf r}}
\newcommand{\bp}{{\bf p}}
\newcommand{\bkk}{{\bf k}}
\newcommand{\bg}{{\bf g}}
\newcommand{\bt}{{\bf t}}
\newcommand{\bu}{{\bf u}}
\newcommand{\bq}{{\bf q}}
\newcommand{\bG}{{\bf G}}
\newcommand{\bP}{{\bf P}}
\newcommand{\bJ}{{\bf J}}
\newcommand{\bK}{{\bf K}}
\newcommand{\bL}{{\bf L}}
\newcommand{\bR}{{\bf R}}
\newcommand{\bS}{{\bf S}}
\newcommand{\bT}{{\bf T}}
\newcommand{\bQ}{{\bf Q}}
\newcommand{\bA}{{\bf A}}
\newcommand{\bH}{{\bf H}}
\newcommand{\btG}{\tilde{{\bf G}}}

\newcommand{\braket}[2]{\left\langle #1 | #2 \right\rangle}
\newcommand{\mel}[3]{\left\langle #1 \left| #2 \right| #3 \right\rangle}

\newcommand{\bdel}{\boldsymbol{\delta}}
\newcommand{\bsig}{\boldsymbol{\sigma}}
\newcommand{\beps}{\boldsymbol{\epsilon}}
\newcommand{\bnu}{\boldsymbol{\nu}}

\newcommand{\bgt}{\tilde{\bf g}}

\newcommand{\brh}{\hat{\bf r}}
\newcommand{\bph}{\hat{\bf p}}

\newcommand{\ga}{\alpha}
\newcommand{\gm}{\mu}
\newcommand{\gb}{\beta}
\newcommand{\gd}{\delta}
\newcommand{\ep}{\epsilon}
\newcommand{\gl}{\lambda}
\newcommand{\go}{\omega}

\newcommand{\bra}[1]{\left\langle #1 \right |}
\newcommand{\ket}[1]{\left| #1 \right\rangle}
\newcommand{\bk}[2]{\left\langle #1 | #2 \right\rangle}

\newcommand{\vect}[1]{\boldsymbol{\mathrm{#1}}}
\newcommand{\op}[1]{\hat{#1}}
\newcommand{\e}[1]{\mathrm{e}^{#1}}

\author{N. Ray$^{1}$}
\author{M. Fleischmann$^{1}$}
\author{D. Weckbecker$^{1}$}
\author{S. Sharma$^{2}$}
\author{O. Pankratov$^{1}$}
\author{S. Shallcross$^{1}$}
\email{sam.shallcross@fau.de}
\affiliation{1 Lehrstuhl f\"ur Theoretische Festk\"orperphysik, Universit\"at Erlangen-N\"urnberg, Staudtstr. 7-B2, 91058 Erlangen, Germany,}
\affiliation{2 Max-Planck-Institut f\"ur Mikrostrukturphysik, Weinberg 2, 06120 Halle, Germany.}

\title{Electron-phonon scattering and in-plane electric conductivity in twisted bilayer graphene}
\date{\today}

\begin{abstract}

We have surveyed the in-plane transport properties of the graphene twist bilayer using (i) a low-energy effective Hamiltonian for the underlying electronic structure, (ii) an isotropic elastic phonon model, and (iii) the linear Boltzmann equation for elastic electron-phonon scattering. We find that transport in the twist bilayer is profoundly sensitive to the rotation angle of the constituent layers. Similar to the electronic structure of the twist bilayer the transport is qualitatively different in three distinct angle regimes. At large angles ($\theta > \,\approx\!\!10^\circ$) and at temperatures below an interlayer Bloch-Gr\"uneisen temperature of $\approx 10$~K the conductivity is independent of the twist angle i.e. the layers are fully decoupled. Above this temperature the layers, even though decoupled in the ground state, are re-coupled by electron-phonon scattering and the transport is different both from single layer graphene as well as the Bernal bilayer. In the small angle regime $\theta <\,\approx\!\!2^\circ$ the conductivity drops by two orders of magnitude and develops a rich energy dependence, reflecting the complexity of the underlying topological changes (Lifshitz transitions) of the Fermi surface. At intermediate angles the conductivity decreases continuously as the twist angle is reduced, while the energy dependence of the conductivity presents two sharp transitions, that occur at specific angle dependent energies, and that may be related to (i) the well studied van Hove singularity of the twist bilayer and (ii) a Lifshitz transition that occurs when trigonally placed electron pockets decorate the strongly warped Dirac cone. Interestingly, we find that, while the electron-phonon scattering is dominated by layer symmetric flexural phonons in the small angle limit, at large angles, in contrast, it is the layer anti-symmetric flexural mode that is most important. We examine the role of a layer perpendicular electric field finding that it affects the conductivity strongly at low temperatures whereas this effect is washed out by Fermi smearing at room temperatures.

\end{abstract}

\maketitle


\section{Introduction}
\label{intro}

The appearance of graphene in the first decade of this century may now be seen as presaging the emergence of a new class of materials: the low dimensional van der Waals heterostructures\cite{geim13}. Amongst such systems the few layer graphenes play a key role as both the most studied example, as well as a group of materials possessing structural simplicity yet unusually rich electronic properties. In this respect one of the most interesting of the few layer graphenes is the graphene twist bilayer, a system that exhibits a remarkably diverse electronic structure as a function of the rotation of the layers. Twist graphene stacks, and as a prototype the graphene twist bilayer, have thus attracted sustained theoretical and experimental attention\cite{shall16,stau16,sboychakov15,son15,sch14,hav14,sar14,yan14,yin14,dob14,beech14,gon13,tab13,lan13,jor13,zou13,shall13,lan13,kim13,tej12,lai12,san12,mel12,yan12,oh12,moon12,bist11,lu11,kin11,mel11,li10,tram10,mor10,mill10,sch10,mel10,shall10,varch08,shall08,shall08a,cam07,lop07,PhysRevB.88.241107}.

From the band theory point of view the twist bilayer is a material both technically as well as conceptually challenging. There are two reasons for this. Firstly, the size of unit cell diverges in the small angle limit $\theta\to 0$, and thus a numerical solution of the band structure problem is of increasing technical difficulty as the twist angle is reduced. Secondly, while all physical properties of the bilayer must be determined by the mutual rotation of the two layers, this angle does not uniquely determine the lattice structure. Thus the usual paradigm of back-folding bands to a superstructure Brillouin zone (BZ) cannot be applied: the twist bilayer does \emph{not have a well defined Brillouin zone}. Instead the system is endowed with an emergent momentum scale, a ``moir\'e momentum'', which depends only on the twist angle and determines an \emph{effective} Brillouin zone, in general different from the \emph{geometric} Brillouin zone\cite{shall13,shall16}. Once the existence of an effective BZ is established, the usual band physics of superstructures follows: single layer states of the constituent layers are back-folded to this effective BZ, leading to a hybridization of these states and the concomitant creation of an angle dependent series of van Hove singularities (vHS), observed both in many theory calculations as well as experiments\cite{yan14,hav14,yan12,oh12,brihuega12,lu11,li10}. In the small angle limit a multitude of these vHS accumulate at the Dirac point, leading to a profoundly complex band structure consisting of a plethora of very high effective mass bands near the Dirac point. The Fermiology of the small angle limit exhibits a corresponding richness, with multiple Lifshitz transitions found in a very small energy window near the Dirac point\cite{shall16}.

This electronic structure is suggestive of correspondingly rich transport properties. A Fermi surface topology that changes dramatically as a function of energy \cite{shall16,sboychakov15}, in combination with the fact that the Fermi energy is a parameter that, experimentally, can be controlled via doping, implies that the twist bilayer is a material for which the transport properties are of great interest. However, thus far experiments have been performed only for the large angle limit\cite{sch10,sanchez12,sch14}, and theoretical calculations restricted to the case of interlayer transport\cite{habib12,perebeinos12,bistritzer10}. The purpose of the present paper, therefore, is to present a systematic investigation of the in-plane transport of the twist bilayer for the complete range of twist angles $1^\circ<\theta<30^\circ$.

A fundamentally important point of interest is the electron momentum transfer due to scattering from, for example, phonons or defects. This is related to the key role of interlayer momentum conservation for the electronic properties of the ground state: electron states from the constituent layers with momentum $\bkk_{1,2}$ (the subscript labels the layer index) scatter only if the condition $\bkk_2-\bkk_1 = \bG_1-\bG_2$ is met ($\bG_{1,2}$ are reciprocal space vectors from each layer). The fact that the reciprocal lattice vectors of the two layers are mutually rotated with respect to each other renders this condition non-trivial, and leads to the emergence of a selection rule for single layer states governed by the ``moir\'e momentum'' $g^{(c)} = [8\pi/(\sqrt{3}a)] \sin\theta/2$. However, with the presence of phonons that provide a momentum $\bq$, the interlayer momentum conservation condition now becomes $\bkk_2-\bkk_1 + \bq = \bG_1-\bG_2$, evidently allowing for many scattering processes that, without the phonon momentum, would be forbidden. In fact, the importance of phonons for understanding the transport properties of the twist bilayer has already been observed in the case of interlayer transport where, based on an effective Hamiltonian theory, it was predicted that the interlayer conductivity should depend sensitively on rotation angle, taking substantial values only near commensurate rotation angles \cite{bistritzer10}. However, once phonons are included into the transport calculation, this dramatic angle dependence of the conductivity is removed\cite{perebeinos12}.

We will explore the conductivity properties of the twist bilayer within the Boltzmann approach where the electron-phonon scattering is very naturally included. The diverging \emph{minimum} unit cell size in the small angle limit $N = (2\sin^2\theta)^{-1}$ with $\theta = \cos^{-1}[(3q^2-1)/(3q^2+1)]$, $q \in \mathbb{N}$ necessitates the use of an effective Hamiltonian approach, in which the interaction of the twisted layers is represented by a continuous ``moir\'e field''\cite{shall16,bist11}, an approach that, in the small angle limit, has been shown to yield excellent agreement with tight-binding calculations\cite{shall16}.

We find that, similar to the ground state electronic structure of the bilayer, the transport properties are qualitatively different in three distinct angle regimes. At large angles $\theta > \,\approx\!\!10^\circ$ the transport may be characterized by an interlayer Bloch-Gr\"uneisen temperature, below which the phonon bath does not possess momenta sufficient to scatter between the cones of the two mutually rotated layers that are separated by a momentum $\Delta K = 8\pi/(3a) \sin\theta/2$ in reciprocal space. Above this temperature, which is rather low at $\approx 10$~K, the two layers, even though decoupled in the ground state, are re-coupled by electron-phonon scattering leading to an in-plane transport different from both of that of single layer graphene as well as any ``simple stacking'' of the bilayer, such as the Bernal stacked bilayer. At intermediate angles we find the energy dependence of the conductivity shows two sharp transitions that may be linked to underlying topological changes in the Fermi surface of the bilayer. One of these is the well known low energy van Hove singularity of the twist bilayer, that leads to a pronounced drop in conductivity associated with the low band velocity saddle point of the van Hove singularity, while the second is driven by the creation of low energy electron pockets decorating the trigonally warped Dirac cones of the twist bilayer. Finally, the small angle regime is associated with a pronounced (almost two orders of magnitude) reduction in conductivity as compared to the large angle ($\theta > 15^\circ$) regime with an energy dependence that exhibits a very complex structure driven by the multiple topological changes in Fermi surface that characterize the bilayer Fermiology at low twist angles.


\section{Model}
\label{model}


\subsection{Electronic structure}
\label{mod-tbl}

In this section we summarize the geometry of the pristine graphene twist bilayer and describe how we calculate its electronic eigenstates $|\vect p \rangle$. The method we use is described in Ref.~\onlinecite{shall16} and we recall it here to introduce our notation in a transparent way, and also because we will refer to the undistorted case repeatedly when solving the more general case of a distorted twist bilayer.

The single layer graphene (SLG) lattice is characterized by the primitive vectors $\vect a_1 = a (1,0)$ and $\vect a_2 = a (\frac{1}{2},\frac{\sqrt{3}}{2})$. With the lattice vectors $\vect R$ being integer combinations of these primitive vectors, the atoms of the $A$- ($B$-) sublattice are found at the positions $\vect R + \vect \nu_{A(B)}$. Here, $\vect \nu_A = (0,0)$ and $\vect \nu_B = \frac{2}{3}(\vect a_1+\vect a_2)$. The reciprocal primitive vectors are $\vect b_1 = \frac{2\pi}{a} (1,-\frac{1}{\sqrt{3}})$ and $\vect b_2 = \frac{2\pi}{a} (0,\frac{2}{\sqrt{3}})$, and integer linear combinations of these vectors give the reciprocal lattice vectors $\vect G$.

A twist bilayer consists of two SLG lattices, which we label with an index $\lambda = \pm 1$, separated by a distance $c$ and with a relative rotation angle $\theta$. We choose the coordinate system such that the graphene layers lie perpendicular to the $z$-axis at $z = \lambda c/2$ and each layer is rotated by an angle $\lambda
\theta/2$ around the $z$-axis. The (real space and reciprocal) vectors of the rotated layers are
denoted with a superscript $[\lambda]$ and related to the unrotated vectors by,
for instance, $\vect a_1^{[\lambda]} = \op{R}_{\lambda \theta /2} \vect a_1$, where
$\op{R}_{\phi}$ is the rotation matrix with rotation angle $\phi$.

\begin{figure}[tbp]
  \centering
  \includegraphics[width=0.98\linewidth]{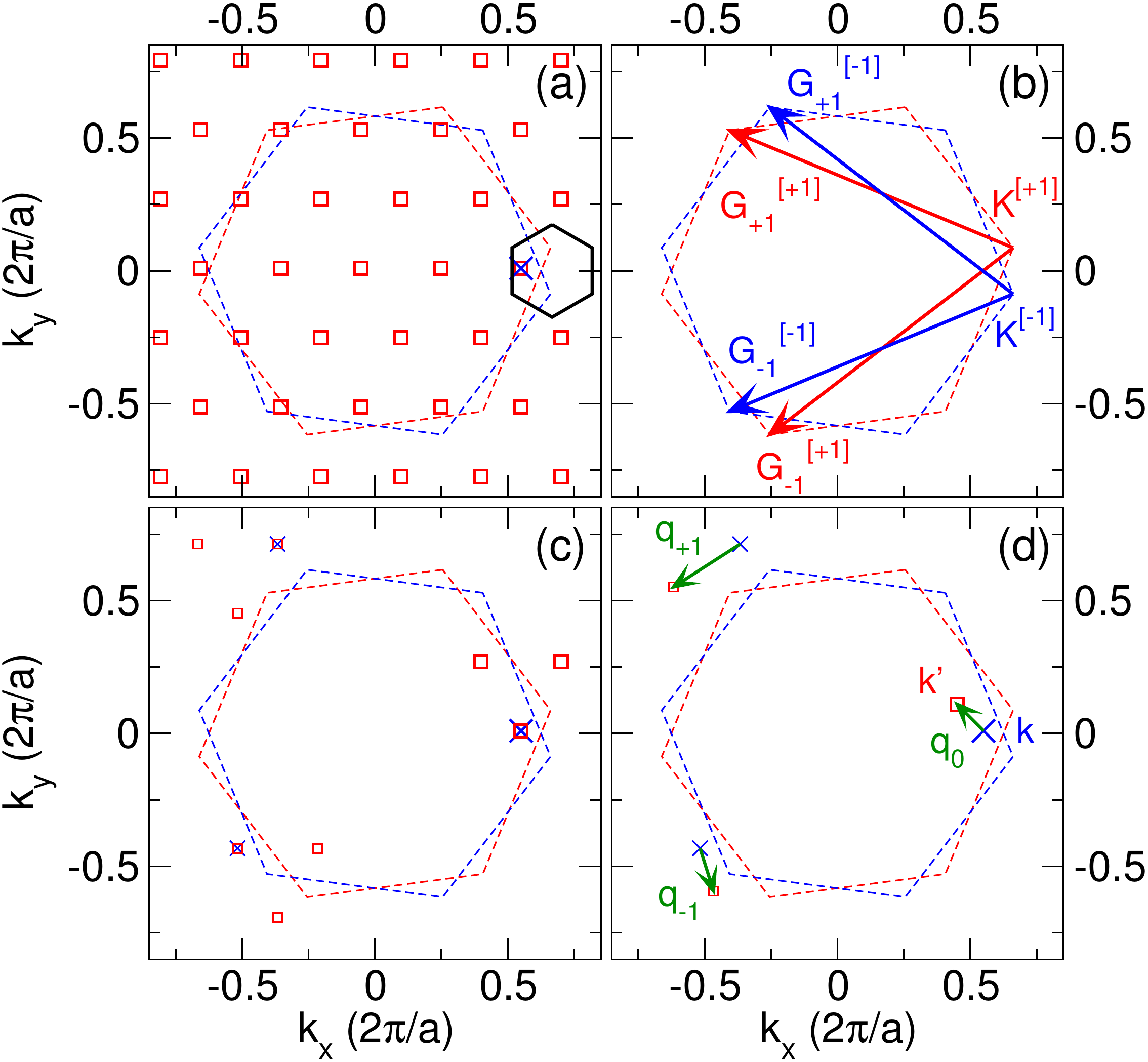}
  \caption{The Brillouin zones of the two rotated single layers with rotation
    angle $\theta = 15^{\circ}$. The red squares
    in panel~(a) depict all states $\vect k'$ of layer $\mu = +1$ that are
    coupled to a given state $\vect k$ (blue X) in layer $\lambda = -1$,
    according to the coupling condition (\ref{eq:couplcond0}). The black hexagon depicts the reciprocal twist bilayer unit cell defined by these coupling vectors, here centered at the unrotated $K$-point. Panel~(b) displays
    the reciprocal SLG lattice vectors $\vect G_j^{[\lambda]}$ that are used in the first star
    approximation; $\lambda = \pm 1$ is the layer index and $j = 0,\pm 1$
    labels the three equivalent $K$-points. The vectors $\vect G_0^{[\lambda]}
    = \vect 0$ are not visible. The first star approximation restricts the 
    interlayer hopping from a given state $\vect k$ to only three states
    $\vect k' = \vect k + \vect G_j^{[\lambda]} - \vect G_j^{[\mu]}$ which is depicted in
    panel~(c) for $\vect k$ (blue cross) in layer $\lambda = -1$ and
    $\vect k'$ (red squares) in layer $\mu = +1$. The smaller symbols depict
    the vectors $\vect k + \vect G_j^{[\lambda]}$
    and $\vect k' + \vect G_j^{[\mu]}$, presented to show that the three allowed hopping sites $\vect k'$ derive from the three equivalent
    $K$-points. With the support of phonons (panel~(d)), interlayer scattering
    from state $\vect k$ to \emph{any} state $\vect k'$ is possible via three
    different processes $j = 0,\pm 1$ with corresponding phonon vectors
    $\vect q_j$, see also Eq.~(\ref{eq:couplcond1star}) of the text. We show two states $\vect k$
    (blue X) and $\vect k'$ (red square) and -- with the smaller symbols -- how the three phonon wave
    vectors $\vect q_j$ derive from the vectors $\vect k + \vect
    G_j^{[\lambda]}$ and $\vect k' + \vect G_j^{[\mu]}$.}
  \label{fig:bz.pdf}
\end{figure}

We wish to solve the Schr\"odinger equation
\begin{align}
\label{eq:schroedinger}
  H_0 |\vect p \rangle = \epsilon_{\vect p} |\vect p\rangle
\end{align}
where $H_0$ represents the twist bilayer Hamiltonian. We approach this problem by expanding the twist bilayer eigenstates $|\vect p \rangle = \sum_{\vect k,\alpha,\lambda} c_{\vect k,\alpha,\lambda}^{\vect p} | \vect k , \alpha,\lambda \rangle$ in a basis of SLG Bloch states
\begin{align}
  | \vect k, \alpha,\lambda \rangle &= \frac{1}{\sqrt{N}}\sum_{\vect R} \e{-i
  \vect k (\vect R^{[\lambda]} + \vect \nu_\alpha^{[\lambda]})} | \vect R^{[\lambda]} + \vect \nu_\alpha^{[\lambda]},\lambda \rangle,
\label{eq:basis00}
\end{align}
that evidently take finite amplitude only on sublattice $\alpha$ of layer $\lambda$. Here $N$ is the number of unit cells in the sample and $| \vect r,\lambda \rangle$ denotes an electron located at the two dimensional position vector $\vect r$ in layer $\lambda$. In this basis $H_0$ is a matrix consisting of matrix elements
\begin{align}
\label{eq:matel}
    \langle \vect k' &, \beta,\mu | H_0 | \vect k, \alpha,\lambda \rangle = \notag \\
&= \frac{1}{N}
\sum_{\vect R, \vect R'} \e{i
  \vect k' (\vect R'^{[\mu]} + \vect \nu_\beta^{[\mu]})} \e{-i \vect k (\vect R^{[\lambda]} + \vect
    \nu_\alpha^{[\lambda]})} \times \notag \\
    &\ \ \ \times \langle \vect R'^{[\mu]} + \vect \nu_\beta^{[\mu]},\mu | H_0 | \vect R^{[\lambda]} + \vect \nu_\alpha^{[\lambda]},\lambda \rangle.
\end{align}
To make further progress we assume that the hopping energy between two sites is a function of their
separation, i.e.
\begin{align}
  \label{eq:t0}
  \langle \vect r+\vect \delta,\mu | H_0 | \vect r,\lambda \rangle =
  t_{\mu,\lambda}(\vect \delta),
\end{align}
where the hopping function $t_{\mu,\lambda}(\vect \delta)$ may be different for interlayer hopping ($\mu = -\lambda$) as compared to intralayer hopping ($\mu = \lambda$) and the argument $\vect \delta$ is the distance between the hopping sites projected on the $xy$-plane. Inserting the Fourier transform $ t_{\mu,\lambda}( \vect \delta ) = (2 \pi)^{-2} \int d^2  q'\  t_{\mu,\lambda} ( \vect q' ) \ \e{- i \vect q' \vect \delta}$ of the hopping function into the matrix element (\ref{eq:matel}), and using lattice vector relations derived from $\sum_{\vect R} \e{i \vect k \vect R } = V_{BZ} \sum_{\vect G} \delta(\vect k - \vect G)$ we find
\begin{align}
\label{eq:matelres0}
\langle \vect k' &, \beta,\mu | H_0 | \vect k, \alpha,\lambda \rangle = \notag \\ 
 &= V_{\text{uc}}^{-1}
   \sum_{\vect G,\vect G'} t_{\mu,\lambda}(\vect
   k' + \vect G'^{[\mu]}) \times \notag \\ 
& \ \ \ \times \e{ i  \vect G \vect
    \nu_\alpha}\e{ -i \vect G' \vect \nu_\beta } \delta_{ \vect
     k + \vect G^{[\lambda]};\vect k'  + \vect G'^{[\mu]}},
\end{align}
where $V_{\text{uc}}$ is the area of the SLG unit cell and the sum runs over all reciprocal SLG lattice vectors $\vect G$, $\vect G'$. From this result we see that two states $|\vect k,\lambda
\rangle$, $|\vect k',\mu \rangle$ are coupled only when the quasi-momenta difference of the two Bloch states satisfies
\begin{align} 
  \label{eq:couplcond0}
  \vect k' - \vect k = \vect G^{[\lambda]} - \vect G'^{[\mu]},
\end{align}
the interlayer conservation of quasi-momentum\cite{shall10,shall13,shall16}, see also Fig.~\ref{fig:bz.pdf}(a).

To solve the Schr\"odinger equation, Eq.~(\ref{eq:schroedinger}), with the help of
the matrix elements Eq.~(\ref{eq:matelres0}) we must choose an explicit
functional form for $t_{\mu,\lambda}(\vect q')$. In this work we apply a Gaussian
function for the real-space hopping energy which, together with the resulting Fourier
transform, we write as
\begin{align}
  \label{eq:gauss0}
  t_{\mu,\lambda}(\vect \delta) &= A_{\mu \lambda}\e{-B_{\mu \lambda}\left(\delta^2+c_{\mu,\lambda}^2\right)},\\
  \label{eq:gauss0ft}
  t_{\mu,\lambda}(\vect q')
&= \frac{ \pi A_{\mu \lambda} \e{-B_{\mu \lambda} c_{\mu,\lambda}^2}}{B_{ \mu
      \lambda}} \ \e{-q'^2/(4B_{ \mu \lambda})},
\end{align}
with constants $A_{\mu \lambda}$ and $B_{\mu \lambda}$ and the $z$-distance between the hopping sites
$c_{\mu,\lambda} = (\mu-\lambda)\cdot c/2$. Upon insertion of Eq.~(\ref{eq:gauss0ft}) into Eq.~(\ref{eq:matelres0}) we then have an explicit expression for obtaining the matrix elements of $H_0$.

For a practical numerical scheme we must truncate the infinite basis of SLG Bloch states and the form of Eq.~(\ref{eq:matelres0}), in conjunction with the exponential decay of the hopping in reciprocal space $t_{\mu,\lambda}(\vect k' + \vect G'^{[\mu]})$, provides a natural cutoff provided $\vect k'$ is close to a high symmetry $K$ point. In fact, for realistic values of the tight-binding constants we require only the vectors $\vect k' + \vect G'^{[\mu]}$ of smallest magnitude which occur when $\vect k' + \vect G'^{[\mu]}$ lies close to one of the first star of special $K$ points, see Fig.~\ref{fig:bz.pdf}(b). If $\vect k'$ is close to $\vect K = \frac{2\pi}{a}(2/3,0)$, this occurs for
$\vect G_0 := \vect 0$, $\vect G_1 := -\vect b_1$, and $\vect G_{-1} := -\vect b_1 - \vect b_2$ and, neglecting all but these three vectors in the summation over reciprocal lattice vectors, Eq.~(\ref{eq:matelres0}) becomes 
\begin{align}
\label{eq:matelres0app}
\langle \vect k' &, \beta,\mu | H_0 | \vect k, \alpha,\lambda \rangle \approx \notag \\ 
 &\approx V_{\text{uc}}^{-1}
   \sum_{j=0,\pm1} t_{\mu,\lambda}(\vect
   k' + \vect G_j^{[\mu]}) \ M_j^{\beta,\alpha} \ \delta_{ \vect
     k + \vect G_j^{[\lambda]};\vect k'  + \vect G_j^{[\mu]}},
\end{align}
with $M_j^{\beta,\alpha}= \e{i \vect G_j (\vect \nu_\alpha-\vect \nu_\beta)}$ given by
\begin{align}
 M_j  = \left(
  \begin{array}{cc}
    1& \e{i \cdot 2 \pi j/3}  \\
    \e{-i \cdot 2 \pi j/3} & 1  \\
  \end{array}
  \right). 
\end{align}
In this first-star approximation each SLG Bloch state $|\vect k,\alpha,\lambda \rangle$ couples to only three SLG states $|\vect k',\beta,\mu \rangle$, with $\vect k' = \vect k + \vect G_j^{[\lambda]} - \vect G_j^{[\mu]}$ and $j=0,\pm 1$ (see Fig.~\ref{fig:bz.pdf}(c)). Note, that in the case of intralayer hopping ($\mu = \lambda$) the coupling condition is $\vect k' = \vect k$ for all three hopping terms, irrespective of $j$. For interlayer hopping ($\mu = -\lambda$) and $j = \pm1$ the coupling vectors $\vect G_j^{[\lambda]} - \vect G_j^{[\mu]}$ form a new reciprocal basis with a length scale $|\vect G_j^{[\lambda]} - \vect G_j^{[\mu]}| = g = 4\sin(\theta/2) /\sqrt{3} \cdot(2\pi/a)$ that decreases monotonically with rotation angle and corresponds to the real space moir\'e lattice scale of $D = 1/(2\sin(\theta/2))$. This is in contrast to the physically irrelevant real space unit cell size and corresponding reciprocal scale, which do not depend monotonically on rotation angle.

In the small energy limit only electron states $\vect k$, $\vect k'$ near $\vect K$ are relevant and we may further simplify the matrix element by expanding $\vect k$, $\vect k'$ around the $K$-point. Defining the small vector
$\vect \kappa = \vect k - \vect K^{[\lambda]}$ and the three equivalent $K$-points $\vect K_j^{[\lambda]} = \vect K^{[\lambda]} + \vect G_j^{[\lambda]}$ we write the argument of the hopping function in
Eq.~(\ref{eq:matelres0app}) as $\vect k' + \vect G_j^{[\mu]} = \vect k + \vect G_j^{[\lambda]} = \vect \kappa + \vect K_j^{[\lambda]}$. The expansion of Eq.~(\ref{eq:gauss0ft}) up to first order is 
\begin{align}
  t_{\mu,\lambda}(\vect k + \vect G_j^{[\lambda]}) \approx t_{\mu,\lambda}(\vect
  K)\left(1 - \frac{\vect \kappa^{[-\lambda]} \vect K_j}{2B_{\mu \lambda}}\right),
\label{eq:ktaylor}
\end{align}
where we have used the relation $\vect \kappa \vect K_j^{[\lambda]} = \vect \kappa^{[-\lambda]} \vect K_j$ for the scalar product of rotated vectors. To examine inter- and intra-layer coupling separately, we write the matrix $H_0$ in a layer-space block matrix form, i.e., as subdivided into four blocks according to layer index, such that the two blocks on the diagonal represent the intralayer hopping of the individual layers, and the two off-diagonal blocks represent the interlayer hopping between the layers. Let us first consider the intralayer blocks ($\mu = \lambda$). Using the expansion (\ref{eq:ktaylor}) in Eq.~(\ref{eq:matelres0app}), we find
\begin{align}
\langle \vect k'&,\lambda | H_0 | \vect k, \lambda \rangle \approx
\delta_{\vect k,\vect k'} \hbar v_F \times
\notag \\
&\times \left(
\begin{array}{cc}
 \epsilon_0/(\hbar v_F) & \kappa^{[-\lambda]}_x + i \kappa^{[-\lambda]}_y\\
 \kappa^{[-\lambda]}_x - i \kappa^{[-\lambda]}_y &  \epsilon_0/(\hbar v_F)
\end{array}
\right),
\label{eq:matintraapp}
\end{align}
which is the standard SLG Dirac-Weyl Hamiltonian for a rotated coordinate system. Here, the Fermi velocity is $\hbar v_F = - \pi t_{\lambda,\lambda}(\vect K)/(a V_{\text{uc}}B_{\lambda \lambda})$ and the physically irrelevant energy shift $\epsilon_0 = 3 t_{\lambda,\lambda}(\vect K)/V_{\text{uc}}$ may be set to zero. For the interlayer interaction ($\mu = -\lambda$) we retain only the zeroth order of Eq.~({\ref{eq:ktaylor}}) to find the
interlayer matrix elements
\begin{align}
\langle \vect k'&,\beta,-\lambda | H_0 | \vect k,\alpha, \lambda \rangle \approx
\notag \\
 &\approx  V_{\text{uc}}^{-1}\ t_{-\lambda,\lambda}(\vect K)
   \sum_{j=0,\pm1} M_j^{\beta,\alpha} \ \delta_{ \vect
     k + \vect G_j^{[\lambda]};\vect k'  + \vect G_j^{[-\lambda]}},
\label{eq:matinterapp}
\end{align}
which do not depend on the wave vectors $\vect k$, $\vect k'$ except via the coupling condition (\ref{eq:couplcond0}).
Equations (\ref{eq:matintraapp}-\ref{eq:matinterapp}) allow us to set up a compact Hamiltonian $H_0$ that, as has been shown in Ref.~\onlinecite{shall16}, reproduces well the exact tight-binding spectrum within an energy window of $\pm 0.4\,$eV about the Dirac point.


\subsection{Phonons}
\label{mod-phs}

We will now consider arbitrary deformations of the twist bilayer lattice, described by the three-dimensional displacement $\vect u^{(\lambda)}(\vect r)$ of the layer $\lambda$ at the two-dimensional position $\vect r$. In this section, we will derive  the twist bilayer phonon modes from a simple elastic model and finally express the quantity $\vect u^{(\lambda)}(\vect r)$ in terms of the corresponding phonon amplitudes.

We approximate the twisted bilayer lattice by a bilayer of isotropic elastic planes. The resulting phonon spectrum only contains the low energy (quasi-)acoustic phonon modes and is independent of rotation angle. This approximation is valid, because due to the weak interlayer interaction, the stacking order has only little influence on the bilayers vibrational properties.\cite{cocemasov13}

The continuous elastic bilayer has six phonon modes $(\sigma,\nu)$, which are labeled by polarization $\nu\in \{\text{l},\text{t},\text{f}\}$ and by symmetry $\sigma \in \{+,-\}$ with respect to the layer index. For a phonon with wave vector $\vect q$ the polarization can be longitudinal ($\nu = \text{l}$), in-plane transverse ($\nu = \text{t}$) or flexural i.e. normal to the plane ($\nu = \text{f}$), and is given by the direction $\vect e_{\vect q,\nu}$ of the displacement vector, which respectively is  $\vect e_{\vect q,\text{l}} = \hat{\vect q}$, $\vect e_{\vect q,\text{t}} = \hat{\vect z} \times \hat{\vect q}$ and 
$\vect e_{\vect q,\text{f}} = \hat{\vect z}$, where $\hat{\vect q} = \vect q / |\vect q|$ and $\hat{\vect z} =(0,0,1)$. Layer-symmetric (-antisymmetric) phonons are labeled with $\sigma\-\-=\-\-+$ ($\sigma\-\-=\-\--$) and correspond to oscillations in which the layers displace relative to each other in the same or in opposite directions, i.e. $\vect u^{(\lambda)}(\vect r) = \sigma \cdot \vect u^{(-\lambda)} (\vect r)$.

In order to express the displacement vector $\vect u^{(\lambda)}(\vect r)$ in terms of phonon amplitudes $u_{\vect q,\sigma,\nu}$, we Fourier transform the coordinate $\vect r$ to reciprocal space with $\vect u(\vect r) = V/(2 \pi)^2 \int d^2 q \ \vect u_{\vect q} \e{-i \vect q \vect r}$, with $V$ being the two-dimensional volume of the sample. Moreover, we switch from the layer index $\lambda$ to the symmetry-index $\sigma$ via the transformation $\vect u_{\sigma} = (\vect u^{(+1)} + \sigma \vect u^{(-1)})/\sqrt{2}$, and the inverse transformation $\vect u^{(\lambda)} = (\vect u_{+} + \lambda \vect u_{-})/\sqrt{2}$. These transformations may be written in a more compact form as $\vect u_{\sigma} =
\sum_{\lambda}s_{\sigma,\lambda}\vect u^{(\lambda)}/\sqrt{2}$ and  $\vect u^{(\lambda)} = \sum_{\sigma}s_{\lambda,\sigma} \vect u_{\sigma}/\sqrt{2}$, with $s_{\lambda,\sigma} = \lambda^{(1-\sigma)/2} = \pm 1$. Finally we decompose $\vect u_{\vect q,\sigma} = \sum_{\nu}u_{\vect q,\sigma,\nu} \cdot \vect e_{\vect q,\nu}$ into polarization components.

The displacement vector takes the form
\begin{align}
\label{eq:phtrafo}
  \vect u^{(\lambda)}(\vect r) =
  \frac{V}{\sqrt{2}(2 \pi)^2} \sum_{\sigma,\nu} \int
  d^2 q \ u_{\vect q,\sigma,\nu} s_{\lambda,\sigma}\e{-i \vect q \vect r}
  \vect e_{\vect q,\nu}.
\end{align}
The phonon amplitudes can be expressed in terms of phonon-creation and -annihilation operators
\begin{align}
\label{eq:phampl}
  u_{\vect q,\sigma,\nu} = \frac{\hbar}{\sqrt{2 \rho V \omega_{\vect
        q,\sigma,\nu}}} \left( a_{\vect q,\sigma,\nu}^{\dagger} + a_{-\vect
    q,\sigma,\nu}\right),
\end{align}
where  $\rho V$ is the mass of the bilayer sample, $a_{\vect q,\sigma,\nu}^{\dagger}$ ($a_{\vect q,\sigma,\nu}$) creates (annihilates) a phonon of mode $(\sigma,\nu)$ and wave vector $\vect q$, and $\omega_{\vect q,\sigma,\nu}$ is the phonon energy.

\begin{figure}[tbp]
  \centering
  \includegraphics[width=0.98\linewidth]{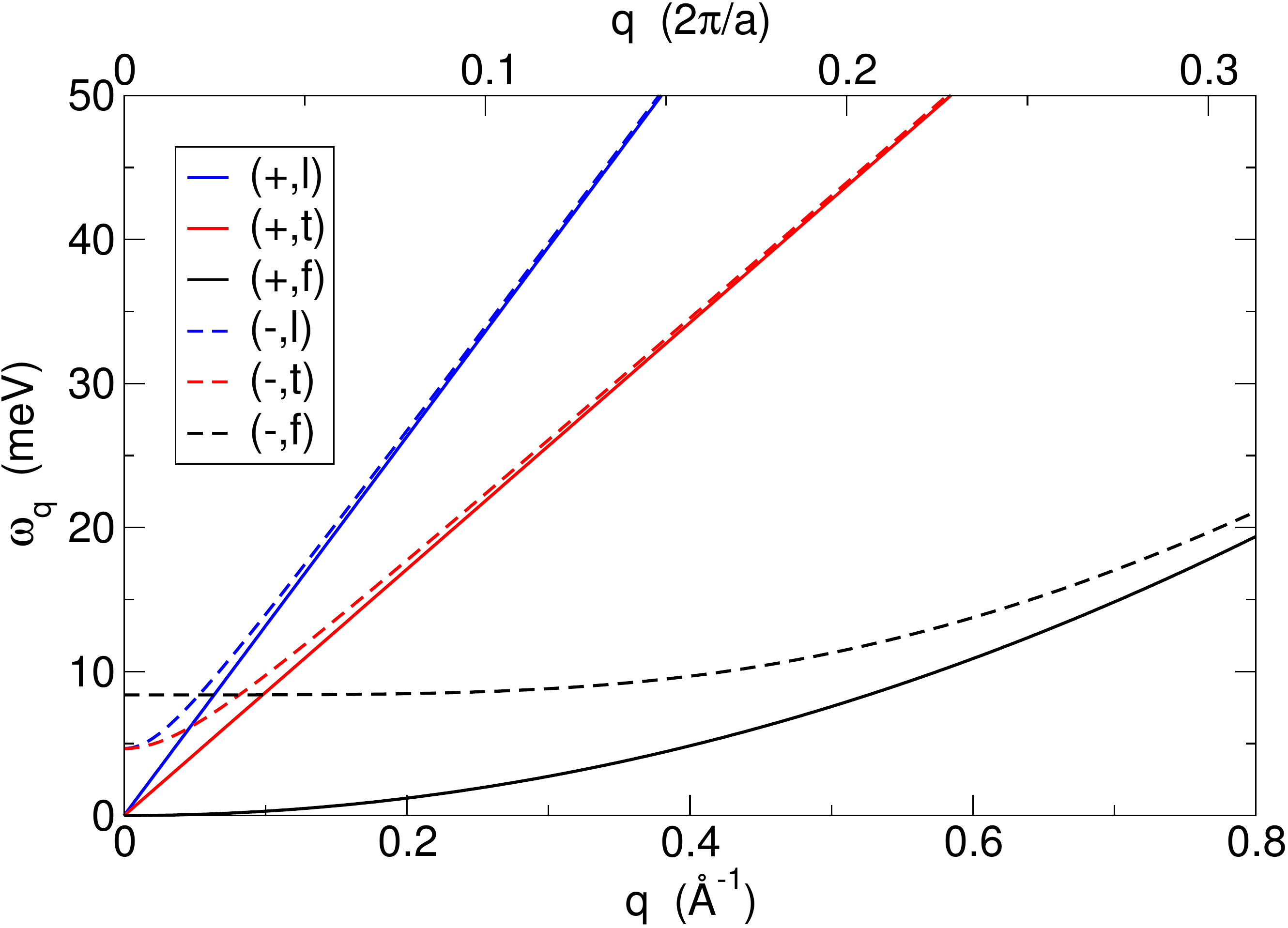}
  \caption{Phonon dispersions of the low energy phonon modes that we consider in this work, see Eq.~(\ref{eq:phdspl}-\ref{eq:phdspf}) for the phonon dispersions and
Table~\ref{tab:parameters} 
for the relevant parameters.}
  \label{fig:phdsp.pdf}
\end{figure}

The energy dispersions, see Fig.~\ref{fig:phdsp.pdf}, of the six phonon modes of an continuous elastic bilayer are\cite{mariani12}
\begin{align}
\omega_{\vect q,\sigma,\text{l}} &= \sqrt{\alpha_{\text{l}}^2 q^2 +
  \Omega_{\text{l}}^2 \delta_{\sigma,-1} },
\label{eq:phdspl}\\
\omega_{\vect q,\sigma,\text{t}} &= \sqrt{\alpha_{\text{t}}^2 q^2 +
  \Omega_{\text{t}}^2 \delta_{\sigma,-1} },
\label{eq:phdspt}\\
\omega_{\vect q,\sigma,\text{f}} &= \sqrt{\alpha_{\text{f}}^2 q^4 +
  \Omega_{\text{f}}^2 \delta_{\sigma,-1} }.
\label{eq:phdspf}
\end{align}
The symmetric phonons $(+,\nu)$ are purely acoustic and have the same dispersion as in a single two-dimensional layer. Note the quadratic dispersion of the flexural mode $(+,\text{f})$. Antisymmetric modes $(-,\nu)$ have a small energy offset $\Omega_{\nu}$ at $\vect q = 0$ due to the interlayer interaction. This adds some optical flavor to the otherwise acoustic vibrations, and we refer to these antisymmetric phonon modes \emph{quasi-acoustic} phonons. In contrast, \emph{optical} phonons -- not treated by a continuum elastic model -- would require out-of phase oscillations of neighboring atoms in the same layer. As the atomic intralayer coupling is much stronger than the interlayer coupling, the energy of an optical phonon at $\vect q =0$ is much higher than the energy offsets $\Omega_\nu$ of the quasi-acoustic modes $(-,\nu)$ and is expected to play almost no role in electron-phonon scattering of the bilayer.

\subsection{Electron-phonon scattering}
\label{mod-eps}

In this section we calculate the scattering matrix element $\langle \vect p',\Phi' | T | \vect p, \Phi \rangle$, that describe scattering from the twist bilayer eigenstate $| \vect p \rangle$ to the twist bilayer eigenstate $|\vect p'\rangle$ while creating or destroying a phonon, with the concomitant change in the phonon population from $|\Phi \rangle$ to $|\Phi' \rangle$. In this expression, $T$ is the phonon-induced scattering potential which we
treat here as a perturbation. 

Firstly we expand the scattering matrix element in the SLG basis Eq.~(\ref{eq:basis00}) yielding 
\begin{align}
  \langle \vect p',\Phi' | &T | \vect p, \Phi \rangle = \langle \Phi' | \cdot \sum_{\vect g,\alpha,\lambda}\sum_{\vect g',\beta,\mu} \Big[c_{\vect g',\beta,\mu}^{\ast \vect p'} c_{\vect g,\alpha,\lambda}^{\vect  p} \notag \\
  &\times \langle \vect p' + \vect g',\beta,\mu | T | \vect p+ \vect
    g,\alpha,\lambda  \rangle \Big] \cdot |\Phi \rangle,    
\label{eq:matelphon}
\end{align}
To calculate the matrix elements $\langle \vect k',\beta,\mu | T | \vect k,\alpha,\lambda \rangle$, we return to Eq.~(\ref{eq:matel}) and replace the Hamiltonian of the unperturbed twist bilayer $H_0$ by the Hamiltonian of a
the distorted twist system $H = H_0 + T$. The deformation of the twist bilayer enters via the hopping function which now does not depend only on the hopping distance $\vect \delta$ but, due to the spatial variation of the deformation, also on the position $\vect r$ in the bilayer at which the hopping takes place. Instead of Eq.~(\ref{eq:t0}) we write
\begin{align}
  \langle \vect r+\vect \delta,\mu | H_0+T | \vect r,\lambda \rangle =
  t_{\mu,\lambda}(\vect r, \vect \delta).
\end{align}
An explicit expression for $t_{\mu,\lambda}(\vect r, \vect \delta)$ will be discussed subsequently. We proceed as in the
undeformed case by sending the hopping to Fourier space via the double transform $t_{\mu,\lambda}(\vect r, \vect \delta) = V/(2 \pi)^4 \int d^2q \int d^2q' t_{\mu,\lambda}(\vect q, \vect q') \e{-i\vect q \vect r} \e{-i \vect q' \vect \delta}$. The first argument $\vect q$ corresponds to the spatial dependence $\vect r$ of the hopping function and hence $\vect q$ is the phonon wave vector. We insert this Fourier transformation into Eq.~(\ref{eq:matel}) finding
\begin{align}
  \langle \vect k' &, \beta,\mu | H_0+T | \vect k, \alpha,\lambda \rangle
  \notag \\
&=
  V_{\text{uc}}^{-1} \sum_{\vect G, \vect G'}
  \ t_{\mu,\lambda}(\vect q, \vect k'+\vect G'^{[\mu]}) \ \e{i \vect G \vect
    \nu_\alpha} \e{-i \vect G'\vect \nu_\beta},
  \label{eq:matelres}
\end{align}
with the phonon vector $\vect q$ fulfilling
\begin{align}
\label{eq:couplcond}
\vect q = (\vect k'+\vect G'^{[\mu]}) - (\vect k + \vect G^{[\lambda]}).
\end{align}
The resulting Eqs.~(\ref{eq:matelres}) and (\ref{eq:couplcond}) should be compared to the matrix element Eq.~(\ref{eq:matelres0}) and the coupling condition Eq.~(\ref{eq:couplcond0}) of the pristine twist bilayer. In contrast to the previous result, now \emph{all} states $\vect k'$ are allowed to couple to a given state $\vect k$ provided the lattice deformation contains a Fourier component $\vect q$.

In the first star approximation Eq.~(\ref{eq:matelres}) becomes
\begin{align}
  \langle \vect k'&, \beta,\mu | H_0+T | \vect k, \alpha,\lambda \rangle \notag \\ 
  & \approx V_{\text{uc}}^{-1} \sum_{j=0,\pm 1} t_{\mu,\lambda}(\vect q_j, \vect k'+\vect G^{[\mu]}_{j})
   \ M_j^{\beta,\alpha},
\label{eq:matelapp}
\end{align}
with the phonon vectors
\begin{align}
\label{eq:couplcond1star}
\vect q_j = (\vect k'+\vect G^{[\mu]}_j) -( \vect k+\vect G^{[\lambda]}_j),
\end{align}
as depicted in Fig.~\ref{fig:bz.pdf}(d). The result Eq.~(\ref{eq:matelapp}) shows that in the first star approximation scattering between any given $\vect k$ and $\vect k'$ is possible via three different processes $j = 0,\pm 1$ with corresponding phonon vectors $\vect q_j$, deriving from the three equivalent $K$-points.

We now consider an explicit hopping function $t_{\mu,\lambda}(\vect r, \vect \delta)$ and its Fourier transform. We use the ansatz from Eq.~(\ref{eq:gauss0}) but due to the deformation $\vect u^{(\lambda)}(\vect r)$ of the lattice the hopping distance between site $\vect r$ in layer $\lambda$ and site $\vect r+\vect \delta$ in layer $\mu$ changes by $\vect u^{(\mu)}(\vect r+\vect \delta) - \vect u^{(\lambda)}(\vect r)$. We decompose this displacement into in-plane and out-of-plane components $\vect u^{(\lambda)}(\vect r) = \bar{\vect u}^{(\lambda)}(\vect r) + h^{(\lambda)}(\vect r) \hat{\vect z}$ and write the real space hopping function as
\begin{align}
  t_{\mu,\lambda}(\vect r, \vect \delta) = &A_{\mu \lambda} \e{-B_{\mu \lambda}
    \left(\vect \delta + \bar{\vect u}^{(\mu)}(\vect r + \vect \delta) -
    \bar{\vect u}^{(\lambda)} (\vect r) \right)^2}  \notag \\
    & \times \e{-B_{\mu \lambda}\left(c_{\mu,\lambda} + h^{(\mu)}(\vect r + \vect \delta) - h^{(\lambda)}(\vect r)\right)^2}.
\end{align}
Note that the $\vect r$ dependence is only due to the deformation field $\vect u^{\lambda}(\vect r)$, as required. Considering only small lattice deformations, we can Taylor expand the hopping function up to linear order finding 
\begin{align}
  \label{eq:taylor2}
  t_{\mu,\lambda}(\vect r, \vect \delta)  \notag \\
  \approx t_{\mu,\lambda}(\vect \delta)\Big[1 &- 2B_{\mu \lambda} \vect \delta
    \cdot \left(\bar{\vect u}^{(\mu)}(\vect r + \vect \delta) - \bar{\vect
    u}^{(\lambda)}(\vect r)\right) \notag \\
 &- 2B_{\mu \lambda} c_{\mu,\lambda} \left(h^{(\mu)}(\vect r + \vect \delta) - h^{(\lambda)}(\vect r)\right)\Big].
\end{align}
Here, the zeroth order term is the hopping function $t_{\mu,\lambda}(\vect \delta)$ of the undeformed twist bilayer, see Eq.~(\ref{eq:gauss0}). It enters the Hamiltonian $H_0$ we treated in section \ref{mod-tbl}. The first order terms represent the change of the hopping energy due to one-phonon scattering processes, while terms beyond first order correspond to multiple phonon scattering processes and the mixing of the phonon modes. In what follows we will treat only one-phonon processes and thus have retained only the first order terms in Eq.~(\ref{eq:taylor2}). 

Note, that in Eq.~\eqref{eq:taylor2} a linear contribution of the flexural deformation $h^{(\mu)}(\vect r + \vect \delta) - h^{(\lambda)}(\vect r)$ appears. This is in contrast to the electron-phonon coupling of single layer graphene, where due to the symmetry with respect to $h(\vect r) \rightarrow -h(\vect r)$ flexural phonons can couple only quadratically, i.e. via two-phonon processes, to the electrons.\cite{castro10,mariani10,mariani08} If the symmetry with respect to the $xy$-plane is broken, linear flexural phonon coupling becomes possible, as has also been discussed for AB-stacked bilayer graphene\cite{mariani12}.

To make further progress write the displacement vector $\vect u^{(\lambda)}(\vect r) = \bar{\vect u}^{(\lambda)}(\vect r) + h^{(\lambda)}(\vect r) \hat{\vect z}$ in terms of the phonon amplitudes, see Eq.~(\ref{eq:phtrafo},\ref{eq:phampl}), and insert the Fourier transform $t_{\mu,\lambda}(\vect q, \vect q') = V^{-1} \int d^2r \int d^2\delta t_{\mu,\lambda}(\vect r, \vect \delta) \ \e{i\vect q \vect r} \e{i \vect q' \vect \delta}$ of the hopping function Eq.~(\ref{eq:taylor2}) into the scattering matrix elements Eq.~(\ref{eq:matelapp}).

The final result of this calculation may be expressed as a sum over all phonon modes
\begin{align}
  \langle \vect k' &, \beta,\mu | T | \vect k,\alpha,\lambda \rangle =
  \notag \\
&= \int d^2q \sum_{\eta} \left( a_{\vect q,\eta}^{\dagger} + a_{-\vect
    q,\eta}\right) w^{\vect k',\beta,\mu;\vect k,\alpha,\lambda}_{\vect
    q,\eta},
\label{matelabrv}
\end{align}
where the vibrational mode is specified with $\eta = (\sigma,\nu)$. Here the terms $w^{\vect k',\beta,\mu;\vect k,\alpha,\lambda}_{\vect q,\eta}$, which are the scattering probabilities for scattering from a SLG state $|\vect k,\alpha,\lambda \rangle$ to another SLG state $|\vect k',\beta,\mu \rangle $ while creating a phonon $(\vect q,\eta)$ or destroying a phonon $(-\vect q,\eta)$, are given by
\begin{align}
&w^{\vect k',\beta,\mu;\vect k,\alpha,\lambda}_{\vect q,(\sigma,\text{l})} = 
     \sum_j \delta ( \vect q;\overline{\vect k}^{\prime}_{j}-\overline{\vect k}_{j})
\frac{D_{\mu,\lambda} M_j^{\beta \alpha}}{\sqrt{\omega_{\vect
          q,\sigma,\text{l}}}} \notag \\
& \ \ \ \times i \Big( s_{\lambda,\sigma}
    \e{-\frac{\overline{\vect k}^{\prime 2}_{j}}{4B_{\mu \lambda}}} \overline{\vect
   k}^{\prime}_{j} -  s_{\mu,\sigma}\e{-\frac{\overline{\vect
          k}_{j}^2}{4B_{\mu \lambda}}}
    \overline{\vect k}_{j} \Big) \cdot \hat{\vect{q}},
\label{wdef1} \\
&w^{\vect k',\beta,\mu;\vect k,\alpha,\lambda}_{j; \vect
  q,(\sigma,\text{t})} = 
     \sum_j\delta (\vect q;\overline{\vect k}^{\prime}_{j}- \overline{\vect k}_{j})
    \frac{ D_{\mu,\lambda}
     M_j^{\beta \alpha}}{\sqrt{\omega_{\vect
        q,\sigma,\text{t}}}} \notag \\
& \ \ \ \times i \Big(
    s_{\lambda,\sigma}\e{-\frac{\overline{\vect k}^{\prime 2}_{j}}{4B_{\mu \lambda}}} \overline{\vect
   k}^{\prime}_{j} -  s_{\mu,\sigma} \e{-\frac{\overline{\vect
          k}_{j}^2}{4B_{\mu \lambda}}}
     \overline{\vect k}_{j} \Big) \cdot \left(\hat{\vect{z}} \times
   \hat{\vect{q}}\right),
\label{wdef2} \\
&w^{\vect k',\beta,\mu;\vect k,\alpha,\lambda}_{j;\vect q,(\sigma,\text{f})} =
    \sum_j \delta (\vect q;\overline{\vect k}^{\prime}_{j}- \overline{\vect k}_{j})
    \frac{ D_{\mu,\lambda}
     M_j^{\beta \alpha}}{\sqrt{\omega_{\vect
        q,\sigma,\text{f}}}} \notag \\
& \ \ \ \times 2B_{\mu \lambda}c_{\mu,\lambda}\Big(
    s_{\lambda,\sigma} \e{-\frac{\overline{\vect
          k}^{\prime 2}_{j}}{4B_{\mu \lambda}}} -
      s_{\mu,\sigma}\e{-\frac{\overline{\vect  k}_{j}^2}{4B_{\mu \lambda}}} \Big),
\label{wdef3}
\end{align}
for longitudinal, transverse, and flexural phonons respectively, and where $ D_{\mu \lambda} = \pi \hbar A_{\mu \lambda}\e{-B_{\mu \lambda}c_{\mu,\lambda}^2}/(2 V_{\text{uc}} \sqrt{\rho V}B_{\mu \lambda})$, $\overline{\vect k}_{j} = \vect k+\vect G_j^{[\lambda]}$, $\overline{\vect k}^{\prime}_{j} = \vect k'+\vect G_j^{[\mu]}$ and $s_{\lambda,\sigma} = \lambda^{(1-\sigma)/2}$.

With this result, the matrix element for the phonon induced scattering of twist bilayer states, Eq.~(\ref{eq:matelphon}), may be written as
\begin{align}
  \langle \vect p',&\Phi' | T | \vect p, \Phi \rangle = \notag \\
& =\int d^2q \sum_{\eta}
  \langle \Phi'| a_{\vect q,\eta}^{\dagger} + a_{-\vect q,\eta}  |\Phi
  \rangle W_{\vect q,\eta}^{\vect p';\vect p},
\label{eq:matelpp}
\end{align}
where the term
\begin{align}
W_{\vect q,\eta}^{\vect p';\vect p} = 
\sum_{\vect
  g,\alpha,\lambda}\sum_{\vect g',\beta,\mu} c_{\vect g',\beta,\mu}^{\ast
  \vect p'} c_{\vect g,\alpha,\lambda}^{\vect  p} w^{\vect p'+\vect g',\beta,\mu,\vect p+\vect
  g,\alpha,\lambda}_{\vect q,\eta}
\label{eq:ppscatprob}
\end{align}
is the scattering probability from a twist bilayer state $|\vect p\rangle $ to another twist bilayer state $|\vect p'\rangle$ while creating a phonon $(\vect q,\eta)$ or destroying a phonon $(-\vect q,\eta)$; note that we have $|W_{\vect q,\eta}^{\vect p';\vect p}|^2 = |W_{-\vect q,\eta}^{\vect p;\vect p'}|^2$.

We have derived the scattering matrix elements, Eq.~\eqref{matelabrv}, valid for scattering of low energy states $\vect k$ near $\vect K^{[\lambda]}$ and $\vect k'$ near $\vect K^{[\mu]}$. A similar result can be obtained for scattering in the inequivalent valleys $\vect K'^{[\lambda]}$ and $\vect K'^{[\mu]}$. Within this model we deploy here, however, the $K$ and $K'$ valleys are treated separately, allowing only phonon scattering $K \leftrightarrow K$ and $K'\leftrightarrow K'$. Treating the $K$ and $K'$ valleys independently, however, disregards the fact, that -- via phonon scattering -- interaction between these valleys is possible. For small and intermediate angles $\theta < \approx 15^\circ$ the momentum separation between the $K$ and $K'$ valleys is sufficiently large that only at very high temperatures are these cones connected by phonon scattering. As we are principally interested in low angles and temperatures we will disregard this mechanism. We note however that at large angles this type of scattering mechanism may be important. In particular, note that at $\theta = 30^\circ$ the interlayer interaction of all four valleys will be equally important as the momentum separation of the $K$ cones
\begin{equation}
 \Delta K = \frac{8\pi}{3a}\sin\frac{\theta}{2} 
\end{equation}
and the momentum separation between the $K$ and $K'$ cones
\begin{equation}
 \Delta K = \frac{8\pi}{3a}\sin\left(\frac{\pi}{6}-\frac{\theta}{2}\right)
\end{equation}
is equal. Therefore if a phonon of sufficient momentum exists such that scattering between Dirac cones is possible, then this scattering mechanism will couple all four cones.


\subsection{Calculation of the conductivity}
\label{mod-con}

For the sake of completeness, in this section we describe how the conductivity of the twist bilayer is calculated on the basis of the linearized Boltzmann equation. The conductivity tensor $\sigma$ is defined by $\vect j = \sigma \vect E$, where $\vect E$ is a homogeneous electric field applied to a sample and
\begin{align}
\vect j = -\frac{4}{(2\pi)^{2}} \int d^2p \ e \vect v_{\vect p} f_{\vect p}
\label{eq:jdef}
\end{align}
is the resulting current density. Here, $e$ is the elementary charge, $\vect v_{\vect p} = \nabla_{\vect p} \epsilon_{\vect p} /\hbar$ is the band velocity and $\epsilon_{\vect p}$ is the energy of an twist bilayer eigenstate $|\vect p\rangle$.
The factor 4 accounts for spin degeneracy and equal contributions from the inequivalent $K$- and $K'$-valleys, which we treat independently.
In a steady state, the distribution function $f_{\vect p}$ is constant, and therefore it may be obtained from the Boltzmann transport equation
\begin{align}
\left. \frac{\partial f_{\vect p}}{\partial t}
\right|_{\text{field}}+ \left. \frac{\partial f_{\vect p}}{\partial t} \right|_{\text{scattering}}=0,
\label{eq:boltzmann}
\end{align}
where the field term is
\begin{align}
\left. \frac{\partial f_{\vect p}}{\partial t}
\right|_{\text{field}} = - \frac{\partial f_{\vect p}^0}{\partial
  \epsilon_{\vect p}} e \vect v_{\vect p} \cdot  \vect E,
\label{eq:field}
\end{align}
with the Fermi-Dirac distribution function $f_{\vect p}^0 = (\e{\beta(\epsilon_{\vect p}-\epsilon_F)} + 1)^{-1}$. The scattering term in Eq.~(\ref{eq:boltzmann}) is determined by the Fermi Golden Rule
\begin{align}
 &\left.  \frac{\partial f_{\vect p}}{\partial
    t}\right|_{\text{scattering}} = \notag \\
  &= \phantom{{}}-\phantom{{}} \frac{2 \pi}{\hbar} \frac{V}{(2\pi)^2} \int d^2p'\sum_{\Phi_f} \left|
    \langle \vect p', \Phi_f | T | \vect p, \Phi_i \rangle \right|^2 \times
    \notag \\
& \hspace{0.3\linewidth} \times  f_{\vect p}(1-f_{\vect p'})
    \delta(\epsilon_{\vect p',\Phi_f} - \epsilon_{\vect p,\Phi_i})   \notag \\
& \phantom{{}={}} +\frac{2 \pi}{\hbar} \frac{V}{(2\pi)^2} \int d^2p' \sum_{\Phi_f} \left| \langle \vect p, \Phi_f | T |
    \vect p', \Phi_i \rangle \right|^2 \times \notag \\ 
& \hspace{0.3\linewidth} \times f_{\vect p'}(1-f_{\vect p})\delta(\epsilon_{\vect p,\Phi_f} - \epsilon_{\vect p',\Phi_i}),
\end{align}
where scattering processes between all electron states $|\vect p\rangle, |\vect p'\rangle$ and all phonon configurations  $|\Phi_i\rangle, |\Phi_f\rangle$ are
considered and $\epsilon_{\vect p,\Phi} = \epsilon_{\vect p}+\epsilon_{\Phi}$ is the sum of the energies of an electron $|\vect p\rangle$ and the phonon
population $|\Phi\rangle$. Inserting the generic electron-phonon scattering matrix elements from Eq.~(\ref{eq:matelpp}) we obtain
\begin{align}
 &\left.  \frac{\partial f_{\vect p}}{\partial
    t}\right|_{\text{scattering}} = \frac{2 \pi}{\hbar} \int \int
  d^2p' d^2q \sum_{\eta} \left|W^{\vect p';\vect p}_{\vect q,\eta}\right|^2 \notag \\
  & \ \ \times \Big[ \phantom{+} f_{\vect p'}(1-f_{\vect
  p}) (1 + n_{\vect q,\eta})  \delta(\epsilon_{\vect
  p'} - \epsilon_{\vect p} - \omega_{\vect q,\eta}) \notag \\ 
& \ \ \phantom{\times \Big[} + f_{\vect p'}(1-f_{\vect
  p}) n_{\vect q,\eta}  \delta(\epsilon_{\vect
  p'} - \epsilon_{\vect p} + \omega_{\vect q,\eta}) \notag \\
&  \ \ \phantom{\times \Big[}- f_{\vect p}(1-f_{\vect
  p'}) (1 + n_{\vect q,\eta})  \delta(\epsilon_{\vect
  p'} - \epsilon_{\vect p} + \omega_{\vect q,\eta})  \notag \\
& \ \ \phantom{\times \Big[} - f_{\vect p}(1-f_{\vect
  p'}) n_{\vect q,\eta}  \delta(\epsilon_{\vect
  p'} - \epsilon_{\vect p} - \omega_{\vect q,\eta}) \Big],
\end{align}
where $n_{\vect q,\eta}$ is the phonon occupation number and $\omega_{\vect q,\eta}$ is the phonon energy. The four terms in the square bracket represent the four different scattering processes for scattering to or from electron state $|\vect p \rangle$ while creating or annihilating one phonon. We linearize the scattering term
\begin{align}
f_{\vect p} &\approx  f^0_{\vect p} - \frac{\partial
  f^0_{\vect p}}{\partial \epsilon_{\vect p}} \phi_{\vect p}
\label{eq:fermilin}
\end{align}
and assume the phonon occupation $n_{\vect q,\eta}$ is given by the Bose-Einstein distribution $n^0_{\vect q,\eta}  = (\e{\beta \omega_{\vect q,\eta}} - 1)^{-1}$.
Employing the elastic scattering approximation $\omega_{\vect q,\eta} \ll \epsilon_{\vect p}$ and retaining only terms linear in $\phi_{\vect p}$, the Boltzmann transport equation (\ref{eq:boltzmann}) is transformed to 
\begin{align}
\vect v_{\vect
  p} \cdot e \vect E & =  \int d^2p' \ P_{\vect p',\vect p} ( \phi_{\vect p'}  -  \phi_{\vect p}) \cdot \delta(\epsilon_{\vect p'}
-\epsilon_{\vect p}),
\label{eq:linearboltzmann1}
\end{align}
with
\begin{align}
P_{\vect p',\vect p}&= \frac{2 \pi}{\hbar} \int d^2q \sum_{\eta} \left|W^{\vect
  p';\vect p}_{\vect q,\eta}\right|^2 \cdot 2 \omega_{\vect q,\eta}
  \frac{\partial n^0_{\vect q,\eta}}{\partial \omega_{\vect q,\eta}}.
\label{eq:linearboltzmann2}
\end{align}
The above formulae are generic for the linearized Boltzmann transport equation within the elastic approximation. For the specific case of the twist bilayer graphene we insert the appropriate
twist bilayer electron-phonon scattering probabilities $W^{\vect p';\vect p}_{\vect q,\eta}$, as given in Eq.~(\ref{eq:ppscatprob}) and Eqs.~(\ref{wdef1}-\ref{wdef3}), and then solve for the unknown function $\phi_{\vect p}$.

In order to solve Eq.~(\ref{eq:linearboltzmann1}) we transform to a basis of functions $\psi_L(\vect p)$ that are orthonormal on the Fermi surface\cite{allen76}, i.e.
\begin{align}
    \frac{1}{\tilde{N}}\int d^2p \ \psi_L(\vect p)\psi_{L'}(\vect p)
    \delta(\epsilon_{\vect p}-\epsilon_{F}) = \delta_{L,L'},
\label{eq:scalarproduct}
\end{align}
with normalization $\tilde{N} = \int d^2p \ \delta(\epsilon_{\vect p}-\epsilon_{F})$. The basis functions $\psi_L(\vect p)$ may be constructed by Gram-Schmidt orthonormalization from any complete set of basis functions, for example the polynomials of $\vect p$-components $\{1,p_x,p_y,p_x^2,p_xp_y,p_y^2,...\}$. Note, that the orthonormalization (\ref{eq:scalarproduct}) will, in general, yield different sets of functions $\psi_{L}(\vect p)$ for different Fermi energies $\epsilon_F$. All functions depending on $\vect p$, e.g. $\phi_{\vect p}$ and $P_{\vect p',\vect p}$, may then be transformed to the new basis via the relations
\begin{align}
  \phi_L &= \frac{1}{\tilde{N}}\int d^2 p \ \delta(\epsilon_{\vect p} - \epsilon_F) \phi_{\vect p}  \ \psi_L(\vect p), \label{eq:trafo1}\\
  \phi_{\vect p} &= \sum_L \phi_L  \ \psi_L(\vect p), \label{eq:trafo2}\\
  P_{L',L} &= \frac{1}{\tilde{N}^2}\int \int d^2p\ d^2p' \ \delta(\epsilon_{\vect p} -
  \epsilon_F) \delta(\epsilon_{\vect p'} - \epsilon_F) \notag \\ 
  &\ \ \ \ \ \ \ \ \ \ \ \ \ \ \ \ \times P_{\vect p',\vect p}  \ \psi_L(\vect p) \psi_{L'}(\vect p'), \label{eq:trafo3}\\
  P_{\vect p' ,\vect p} &= \sum_{L,L'} P_{L',L}  \ \psi_L(\vect p) \psi_{L'}(\vect
  p'). \label{eq:trafo4}
\end{align}
The linearized Boltzmann equation (\ref{eq:linearboltzmann1}) in the new basis reads
\begin{align}
  \vect v_L \cdot e \vect E &= \tilde{N} \sum_{L'} \left[  P_{L',L} - \sum_{L''}C_{L,L',L''}P_{L'',0}\right]  \phi_{L'},
\label{eq:linboltzl}
\end{align}
with the Clebsh-Gordon coefficients
\begin{align} 
C_{L,L',L''} &= \frac{1}{\tilde{N}} \int d^2p \ \delta(\epsilon_{\vect p} - \epsilon_F)\psi_L(\vect p)\psi_{L'}(\vect p) \psi_{L''}(\vect p),
\end{align}
and where we have chosen the basis function with index $L=0$ to be the constant function $\psi_{0}(\vect p) = 1$. Eq.~(\ref{eq:linboltzl}) is a matrix equation $v_L = \sum_{L'}M_{L,L'}\phi_{L'}$ which can be solved for the unknown vector $\phi_{L'}$ by inversion of the matrix $M_{L,L'}$. Having solved the Boltzmann equation Eq.~(\ref{eq:linboltzl}), we subsequently use Eqs.~(\ref{eq:trafo2}) to calculate the function $\phi_{\vect p}$.  The current density is obtained from Eq.~\eqref{eq:jdef}, which using Eq.~(\ref{eq:fermilin}) we rewrite as
\begin{align}
\vect j = \frac{4}{(2\pi)^{2}} \int d^2p \ e \vect v_{\vect p} \frac{\partial f^0_{\vect p}}{\partial \epsilon_{\vect p}}\phi_{\vect p}.
\label{eq:jdefphi}
\end{align}
Finally we read off the conductivity-tensor $\sigma$ from the equation $\vect j = \sigma \vect E$. Due to Eq.~\eqref{eq:linboltzl}, the proportionality $\vect j \sim |\vect E|$ always holds, i.e. the resulting conductivity is guaranteed to be independent of $|\vect E|$, as required. Note that due to the elastic scattering approximation all scattering processes are between electron states with the same energy and thus different energies can be treated separately in solving the linearized Boltzmann equation. It is only in the last step, the integration of current density Eq.~\eqref{eq:jdefphi}, that different energies contribute according to the Fermi-Dirac
distribution, leading to a conductivity that at Fermi energy $\epsilon_F$ depends on scattering processes of electron states within an energy window $\epsilon_F - k_BT < \epsilon < \epsilon_F + k_BT$.


\section{Computational Details}
\label{comp}

\begin{table}
\caption{\label{tab:parameters} Values of various parameters and physical constants used in the model of section \ref{model}; $a$ is the graphene lattice constant and $c$ the interlayer separation of the bilayer, $A$ and $B$ are parameters that determine the tight-binding matrix elements\cite{lan13}, and $\Omega$ and $\alpha$ parameters that determine the spectrum of the six low energy phonon modes of the bilayer. See text for further details.}
\begin{ruledtabular}
\begin{tabular}{ldl||ldl}
$a=$ & 2.46 & \AA & $\Omega_{\text{l}} = \Omega_{\text{t}}=$ & 4.65 & meV\\
$c=$ & 3.35 & \AA & $\Omega_{\text{f}}=$ & 8.38 & meV \\
\hline
$A_{\lambda,\lambda}=$ & -8.45 & eV & $\alpha_{\text{l}}=$ & 131.64 & $\text{meV} \text{\AA}$ \\
$B_{\lambda,\lambda}=$ & 0.66 & $\text{\AA}^{-2}$ & $\alpha_{\text{t}}=$ & 85.57 & $\text{meV} \text{\AA}$ \\
$A_{-\lambda,\lambda}=$ & 50& eV &$\alpha_{\text{f}}=$ & 30.28 & $\text{meV} \text{\AA}^2$ \\
$B_{-\lambda,\lambda}=$ & 0.44 & $\text{\AA}^{-2}$ & & & \\
\end{tabular}
\end{ruledtabular}
\end{table}

For numerical calculations we parameterize the model outlined in the previous section as indicated in Table~\ref{tab:parameters}. The parameters $A_{\mu,\lambda}$ and $B_{\mu,\lambda}$ for the tight-binding hopping energies (Eqs.~(\ref{eq:gauss0}-\ref{eq:gauss0ft})) have been fitted to results of density functional theory calculations for a dataset of small unit cell structures, including both Bernal and twist bilayers\cite{lan13}. The parameters $\alpha_\nu$ of the phonon spectrum, Eq.~(\ref{eq:phdspl}-\ref{eq:phdspf}), describe the in-plane elastic properties of the twist bilayer and can be obtained directly from the phonon dispersion of graphene or graphite, which has been studied extensively both theoretically\cite{aljishi82,dubay03,wirtz04,mounet05,metlov10} as well as experimentally\cite{saito01,maultzsch04,mohr07}. Comparing the phonon dispersion of graphene and graphite reveals that the in-plane elastic properties $\alpha_\nu$ are robust against different stacking configurations of graphene layers and therefore the same values can be used for the twist bilayer. For the parameters $\Omega_\nu$ describing the interaction between the layers we use values in agreement with Ref.~\onlinecite{borysenko11}. It should be noted however, that in the literature a range of values ($\pm 30 \%$) can be found for this constant\cite{yan08,saha08,cocemasov13}.
Furthermore a small dependence of $\Omega_\nu$ on layer rotation is suggested\cite{cocemasov13}, which is neglected in this work as it does not affect our results noticeably.

When calculating the twist bilayer electronic band structure we expand the twist eigenfunctions in SLG basis functions $|\vect p\rangle = \sum_{\vect g,\alpha,\lambda} c^{\vect p}_{\vect p + \vect g,\alpha,\lambda}|\vect p + \vect g,\alpha,\lambda\rangle$. The number of required basis functions, i.e. the dimension of the Hamiltonian that has to be diagonalized at each point in $\vect k$-space, depends on the Fermi energy $\epsilon_F$ and on the rotation angle $\theta$ of the system. In this work we only calculate bilayers with $\theta \geq 1^\circ$ and $|\epsilon| \leq 0.4$ eV, in which case a basis of a few hundred SLG functions has been found to be sufficient\cite{shall13,shall16}.

The computationally most expensive part of the numerical procedure, however, is the calculation of scattering probabilities between all points on the Fermi surface, as the number of such scattering processes $W^{\vect p,\vect p'}$ evidently scales quadratically with $N_{\vect p}$, the number of mesh points on the Fermi surface. Depending on the complexity of the Fermi surface we use a mesh size of $500 < N_{\vect p} < 3000$ points.

Solving the Boltzmann equation requires the inversion of the matrix $M_{L'L} = P_{L',L} - \sum_{L''}C_{L,L',L''}P_{L'',0}$ in Eq.~\eqref{eq:linboltzl}. The dimension of this matrix is given by the number $n_L$ of basis functions $\psi_L(\vect p)$ required to transform the functions $\vect v_{\vect p}$ and $P_{\vect p',\vect p}$ to the basis of orthonormal Fermi surface functions. It turns out that these functions are smooth enough that $n_L = 45$ is sufficient for $\theta \geq 5^\circ$. For smaller rotation angles we increase it gradually to $n_L = 435$ at the smallest rotation angle $\theta = 1^\circ$ that we treat in this work.

With the present model we calculate all four components $\sigma_{ij}$ of the $2\times2$ conductivity tensor. However, in all calculations we find that the conductivity tensor is isotropic, and we can thus write the result as $\sigma_{ij} = \sigma \delta_{ij}$ with $\sigma$ a scalar. In the following we refer to this scalar $\sigma$ when referring to conductivity.


\section{Results}
\label{results}

Having established a formalism within which the conductivity of the twist bilayer may be calculated, in this section we will explore the conductivity over the full range of angles of the twist bilayer. As is by now well known, the twist bilayer exhibits an extraordinary richness of electronic structure as a function of the twist angle (see for example Ref.~\onlinecite{shall16}), and our primary interest here will be (i) to establish the corresponding behaviour for the transport properties and (ii) to relate this transport behaviour to the underlying electronic structure. To this end we will first overview the electronic structure of the twist bilayer.

The ground state electronic structure of the twist bilayer may be characterized by three qualitatively different types of behaviour that occur at three distinct twist angle regimes. At large angles ($\theta >\,\,\approx \!\!15^\circ$) the bilayer is essentially electronically decoupled while, in contrast, at small angles ($\theta <\,\, \approx\!\!2^\circ$) the bilayer is strongly coupled, and exhibits a rich electronic structure that differs significantly both from single layer graphene and any ``simple stacking'' arrangement such as the Bernal (AB) stacked bilayer. These two coupling strength limits are connected by an angle window in which the electronic spectrum is qualitatively that of single layer graphene, but decorated by van Hove singularities occurring due to the intersection of the two Dirac cones from each layer. As the twist angle is reduced these van Hove singularities both increase in number and move continuously towards the Dirac point (reflecting the increasing number of intersection points of the two cones as they move closer together in momentum space). This ends at small angles in the complete destruction of the single layer spectrum.

In order to elucidate the relationship between the underlying electronic structure and the conductivity, we will first set $\partial f_{\vect p}^0/\partial \epsilon_{\vect p} := -\delta(\epsilon_F - \epsilon_{\vect p})$ in Eq.~\eqref{eq:jdefphi}. Under this approximation only states at $\ep = \ep_F$ contribute to the conductivity and this evidently facilitates understanding the relationship between the conductivity $\sigma(\ep_F)$ and the underlying electronic structure. Note that there are two distinct temperature dependencies in the formalism described in Section \ref{model} - a ``Fermionic temperature'' of the electron quasiparticles and a ``Bosonic temperature'' of the phonon bath - and thus this approximation is not a temperature independent approximation.

In the subsequent sections we will successively describe the conductivity at large angles, intermediate angles, and finally in the small angle strong coupling limit.


\subsection{Conductivity at large angles: $\theta > 10^\circ$}
\label{res-con-large}

\begin{figure}[tbp]
  \centering
  \includegraphics[width=0.98\linewidth]{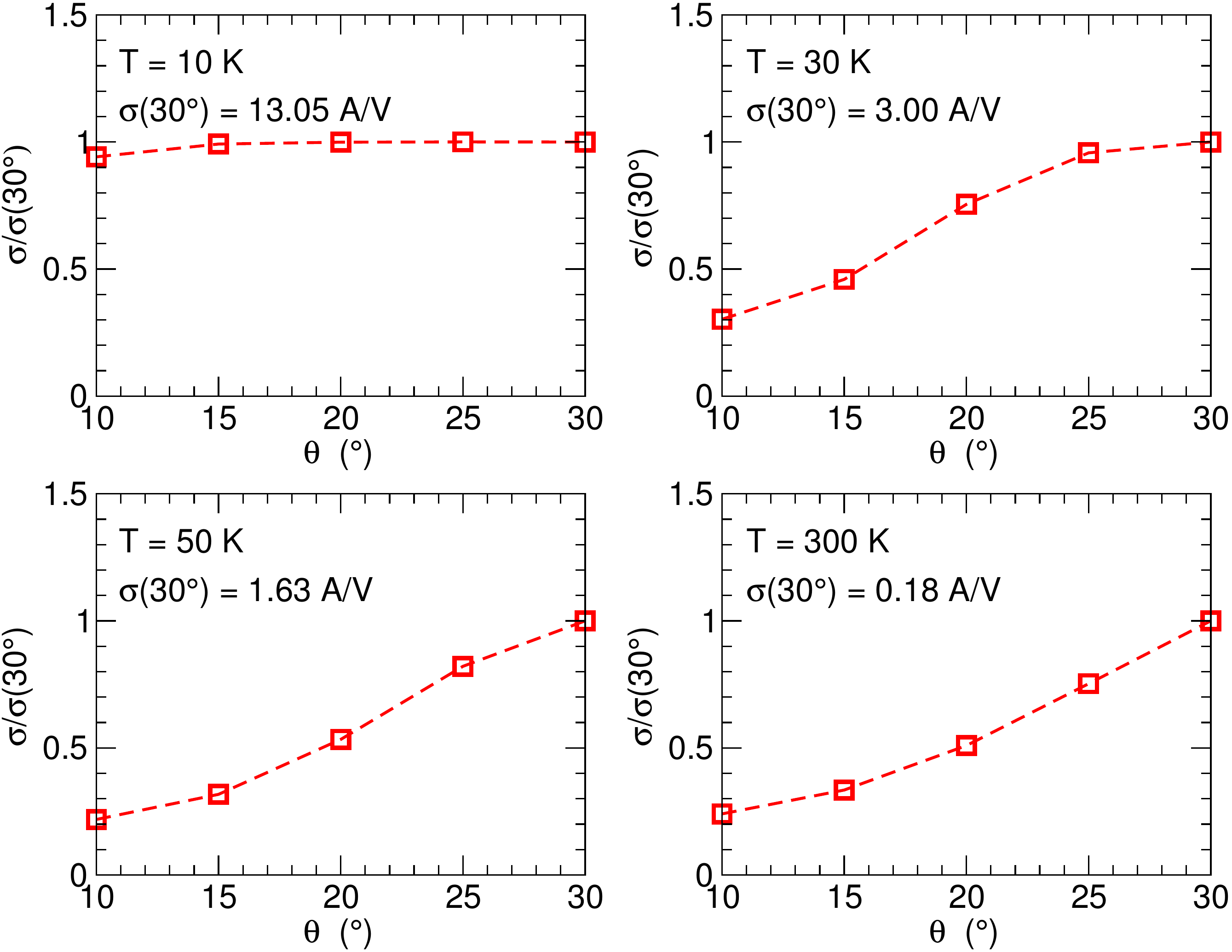}
  \caption{\emph{Transport re-coupling of the bilayer at large angles.} The conductivity $\sigma$ as a function of twist angle for a Fermi energy of $\epsilon_F = 50\,$meV and for temperatures $T = 10$~K to $T=300$~K as indicated. Only at low temperatures does the transport become independent of the twist angle, as may be seen in the $T=10$~K panel for $\theta > 20^\circ$. Thus, in contrast to the ground state electronic structure, only at low temperatures does the conductivity of the twist bilayer become layer decoupled at large angles. The cross-over between the layer decoupled and layer coupled transport behaviour is determined by an angle dependent \emph{interlayer Bloch-Gr\"uneisen temperature}, below which the phonon bath does not possess sufficient momentum to scatter between the two Dirac cones of each layer that are separated in momentum space by $\Delta K = 8\pi/(3a)\sin\theta/2$. Note that in this calculation the temperature dependent Fermi smearing is switched off by setting $\partial f_{\vect p}^0/\partial \epsilon_{\vect p}= -\delta(\epsilon_F - \epsilon_{\vect p})$ in Eq.~\eqref{eq:jdefphi} of the main text, and thus the conductivity reflects only scattering processes occurring at the Fermi energy.
  }
  \label{fig:cond-large}
\end{figure}

\begin{figure}[tbp]
  \centering
  \includegraphics[width=0.98\linewidth]{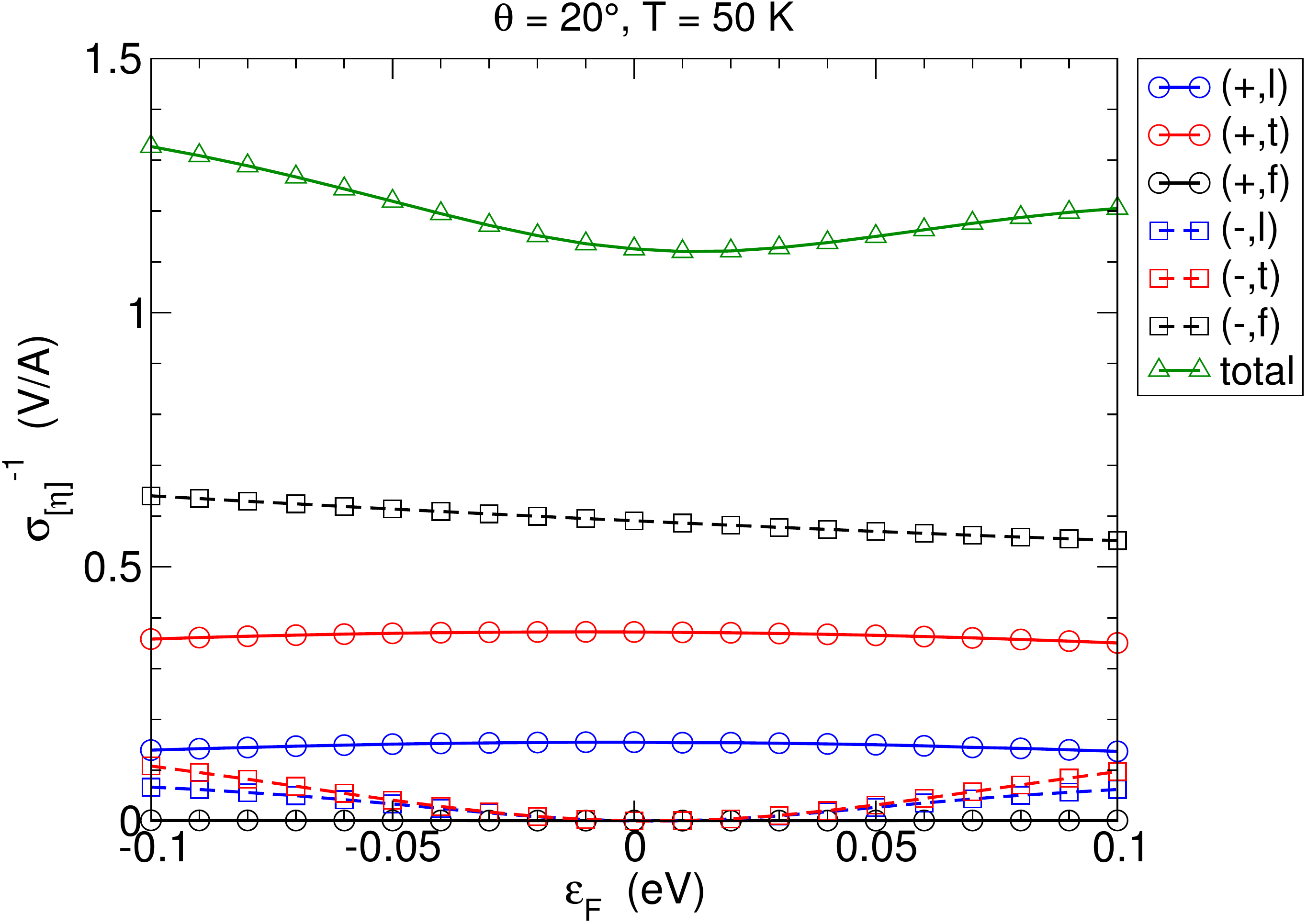}
  \caption{\emph{Particle-hole asymmetry of the transport in the twist bilayer.} Shown is the contribution to the resistivity $\sigma_{[\eta]}^{-1}$ from the 6 distinct phonon modes $\eta=(\sigma,\nu)$ that the bilayer possesses, plotted as a function of the Fermi energy $\epsilon_F$ for temperature $T = 50$~K and a rotation angle of $\theta=20^{\circ}$. In contrast to the case in single layer graphene, or the Bernal stacked bilayer, the transport is asymmetric, a fact that arises as the moir\'e field itself does not possess particle-hole symmetry. Note that the temperature dependent Fermi smearing $\partial f_{\vect p}^0/\partial \epsilon_{\vect p}$ of the electron states is not included, and thus the resulting transport reflects only scattering processes that take place at the Fermi energy.
}
  \label{fig:phmodes-large.pdf}
\end{figure}

In Fig.~\ref{fig:cond-large} we present the conductivity of the twist bilayer for $10^\circ < \theta < 30^\circ$ and four temperatures; $T = 10$~K, $30$~K, $50$~K, and $300$~K. The Fermi energy is $\epsilon_F = 50\,$meV in all four panels (the results do not change qualitatively upon changing the Fermi energy). In dramatic contrast to the ground state electronic structure, which at these energies would be identical to that of single layer graphene and therefore angle independent, we see that the transport properties show a pronounced angle dependence once $T > 10$~K. This can be understood by noting that the Dirac cones from each layer are separated in momentum space by

\begin{equation}
 \Delta K = \frac{8\pi}{3a} \sin\frac{\theta}{2}.
\end{equation}
%
Thus once the phonon bath has sufficient momentum to scatter between these two cones (see Fig.~\ref{fig:phdsp.pdf}) then even though the ground state of the bilayer is electronically decoupled, the transport problem re-couples the layers. We should stress that the purpose of the calculations we present in this section is simply to probe the question of transport coupling of the bilayer. An accurate calculation of in-plane transport in the large angle decoupled limit requires both second order phonon scattering processes, as the in-plane flexural phonon couples at order $O(q^2)$, as well as $K \leftrightarrow K'$ scattering to be included.

To understand the angle-dependent interlayer coupling in more detail, we consider the contributions to the resistivity of the individual phonon modes; flexural, longitudinal, and transverse (we have 6 phonon modes as each of these may be either layer symmetric or layer anti-symmetric in nature). By restricting the sum in Eq.~\eqref{eq:matelpp} to a specific phonon mode $\eta_0$ we obtain the conductivity $\sigma_{[\eta_0]}$ in which only the selected phonon mode is active. As we consider only single electron-phonon scattering events these conductivity contributions fulfill the equation $\sigma^{-1} = \sum_{\eta} (\sigma_{[\eta]})^{-1}$, with $\sigma$ being the conductivity including all phonon modes. As may be seen in Fig.~\ref{fig:phmodes-large.pdf} the anti-symmetric flexural mode $(-,\text{f})$ makes the largest contribution to the resistivity, while in contrast the symmetric flexural phonons $(+,\text{f})$ make a negligible contribution. The anti-symmetric flexural phonon spectrum, as may be seen in Fig.~\ref{fig:phdsp.pdf}, possesses a gap of $\approx 8$~meV, and the magnitude of this gap, along with the momentum separation of the cones $\Delta K$, evidently determines the temperature at which the scattering between the two cones is switched on.

Curiously, we find that the energy dependence of the resistivity does not display particle-hole symmetry, exhibited most strongly by the anti-symmetric flexural mode which has a linear dependence on Fermi energy as may be seen in Fig.~\ref{fig:phmodes-large.pdf}. This is in dramatic contrast to both single layer graphene, as well as the Bernal stacked bilayer, in which the dependence on Fermi energy (or equivalently charge carrier density) is particle-hole symmetric.\cite{mariani12,mariani10} The reason for this lies in the nature of the interlayer coupling, which for the twist bilayer is carried by the position dependent ``moir\'e field'' given by the layer off-diagonal blocks of the twist Hamiltonian, Eq.~\eqref{eq:matinterapp}, a much more complex object than the relatively simple position independent coupling of the AB bilayer. These layer off-diagonal blocks in fact explicitly break particle-hole symmetry, as may be seen by operating with anti-unitary operator $i\sigma_y K$ ($K$ the complex conjugation operator). In contrast the Dirac-Weyl Hamiltonian, as well as the layer off-diagonal blocks of the Bernal bilayer, do possess particle-hole symmetry.


\subsection{Conductivity at intermediate angles: $2^\circ < \theta < 10^\circ$}
\label{res-con-int}

\begin{figure}[tbp]
  \centering
  \includegraphics[width=0.98\linewidth]{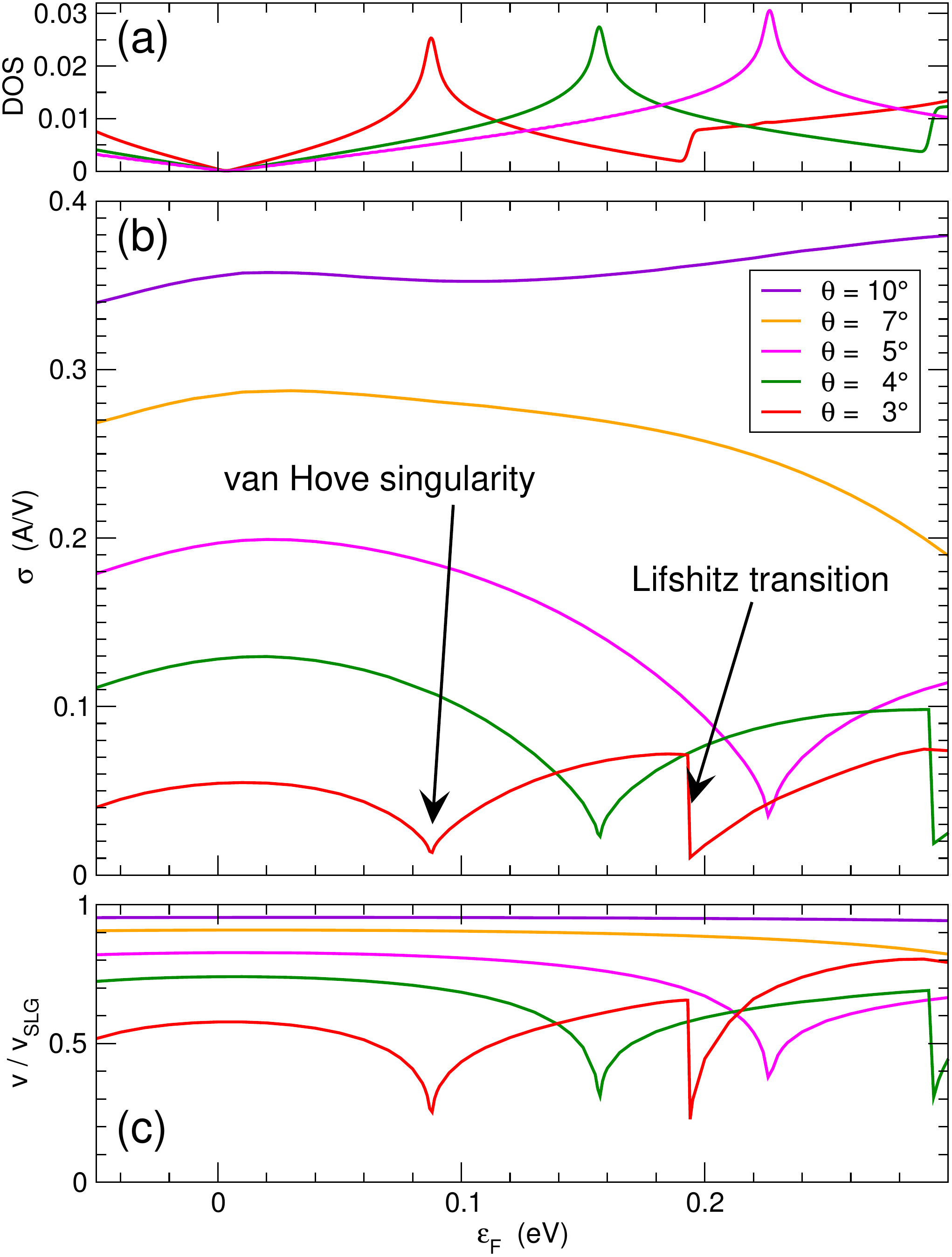}
  \caption{\emph{Band velocity dominated transport regime.} At intermediate angles the transport of the twist bilayer is dominated by band structure effects that result from the hybridization of the two Dirac cones from each layer. To see this compare the band velocity averaged over the Fermi surface $|\vect v|$ (panel (c)) with the conductivity (panel (b)). Clearly, the dramatic changes in conductivity that, for example, occur at 88~meV and 194~meV for the $\theta = 3^\circ$ system are the result of corresponding changes in the Fermi surface averaged band velocity (and similar for all other angles presented). These sudden transitions result from topological changes in the Fermi surface that occur at these energies, and both produce noticeable features in the density of states (see panel (a)). For details and explanation of the changes in Fermi surface topology see Section IV B and Fig.~6. Note that the temperature dependent Fermi smearing is switched off by setting $\partial f_{\vect p}^0/\partial \epsilon_{\vect p}= -\delta(\epsilon_F - \epsilon_{\vect p})$ in Eq.~\eqref{eq:jdefphi} of the main text, and thus the conductivity reflects only scattering processes occurring at the Fermi energy.}
  \label{fig:cond-intermediate}
\end{figure}

At intermediate twist angles the transport behaviour is, as may be seen from Fig.~\ref{fig:cond-intermediate}(b), strikingly different from the large angle case. Shown in Fig.~\ref{fig:cond-intermediate}(b) is the conductivity as a function of energy which, in contrast to the rather smooth dependence seen at large angles, exhibits two points of non-analytic behaviour (recall we have switched off the Fermi smearing). These are labeled by ``van Hove singularity'' and ``Lifshitz transition'' in Fig.~\ref{fig:cond-intermediate}(b) and, within the energy range of the plot, occur for all angles $2^\circ < \theta < 4^\circ$. The former of these is evidently closely connected to the well known van Hove singularities found in the density of states of the twist bilayer at these angles, as may be seen by comparison of the density of states shown in Fig.~\ref{fig:cond-intermediate}(a) and the conductivity plotted in Fig.~\ref{fig:cond-intermediate}(b), while the latter has its origin (as we will shortly show) in a topological change in the Fermi surface, and is responsible for the step like feature seen at density of states at $\approx 0.2$~eV. To investigate these features further we display in Fig.~\ref{fig:cond-intermediate}(c) the band velocity averaged over the Fermi surface $v = \int d^2p |\vect v_{\vect p}| \delta(\epsilon_{F} - \epsilon_{\vect p}) / \int d^2p \delta (\epsilon_{F}-\epsilon_{\vect p})$. Both of the non-analytic features of $\sigma(\epsilon_F)$ can clearly be correlated to similar changes in the average band velocity, implying a band structure origin for both.

\begin{figure}[tbp]
  \centering
  \includegraphics[width=0.98\linewidth]{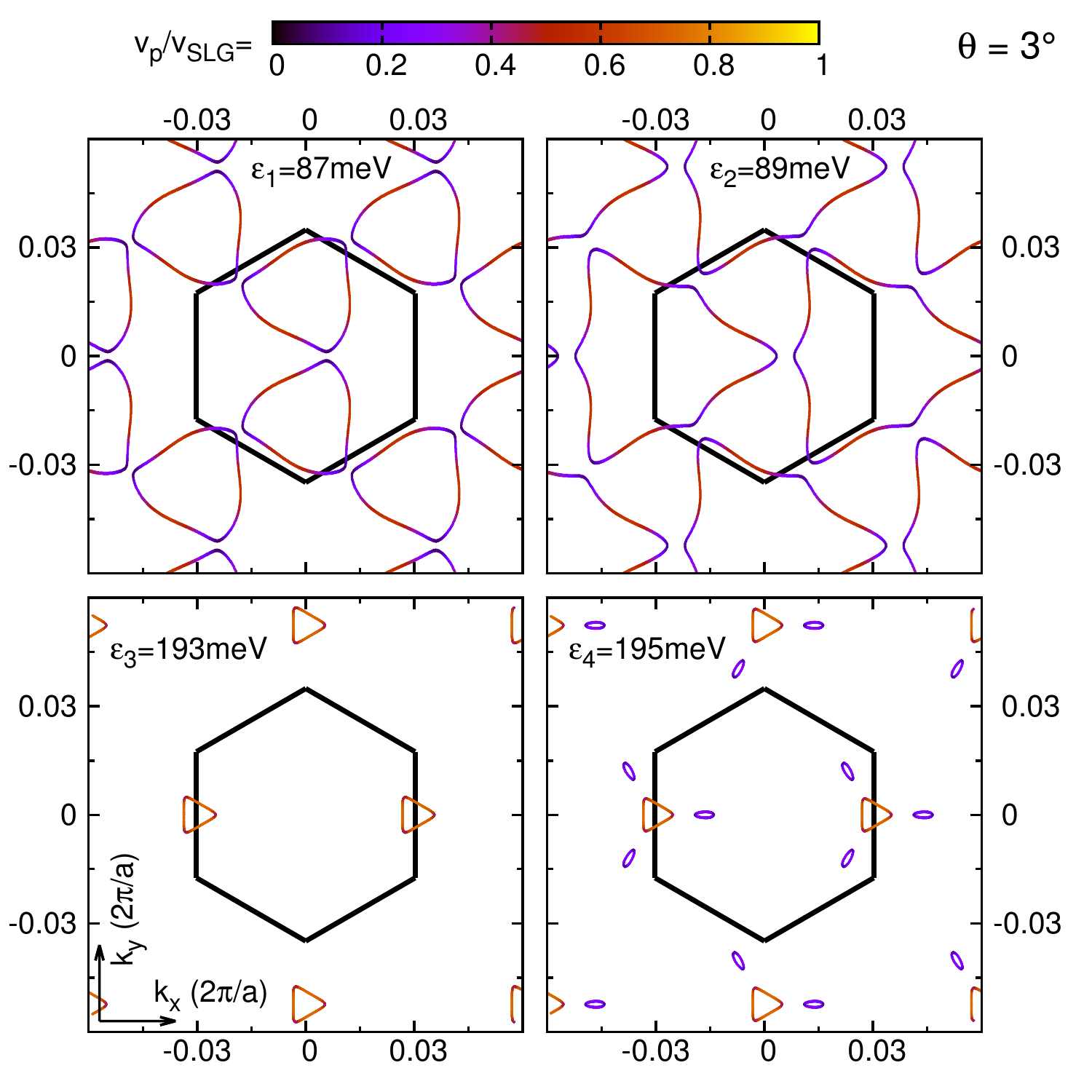}
  \caption{Fermi surfaces of the twist bilayer with rotation angle $\theta = 3^\circ$ shown for four different energies $\epsilon_1 = 87$~meV, $\epsilon_2 = 89$~meV, $\epsilon_3 = 193$~meV, and $\epsilon_4 = 195$~meV. The color encodes the band velocity $|\vect v_{\vect p}|$ in units of the SLG Dirac band velocity $v_{\text{SLG}}$ while the black hexagons depict the twist bilayer reciprocal space unit cell, see also Fig.~\ref{fig:bz.pdf}(a). The Fermi surfaces should be compared to the $\theta = 3^\circ$ conductivity and average band velocity displayed in Fig.~\ref{fig:cond-intermediate} and illustrate the topological changes that underlie the two pronounced features in the conductivity indicated by the arrows in Fig.~\ref{fig:cond-intermediate}.
  }
  \label{fig:fermisf2.pdf}
\end{figure}

In the top two panels of Fig.~\ref{fig:fermisf2.pdf} we present a set of extended Fermi surfaces for the $\theta = 3^\circ$ bilayer and, as indicated, two energies $\epsilon_1 = 87\,$meV and $\epsilon_2 = 89\,$meV that are either side of the point labeled ``van Hove singularity'' in the conductivity found at $\ep = 88\,$meV. As may be seen while at $\epsilon_1$ the Fermi surface consists of two disconnected loops corresponding to the two Dirac cones of each layer, at $\epsilon_2$ these loops have merged to form a qualitatively different Fermi surface. 
This intersection of the cones, and the resulting band repulsion at the intersections, creates a local energy gap that leads both to the van Hove peak in the density of states, as well as the pronounced drop in the Fermi surface averaged band velocity. This drop in average band velocity leads to a corresponding drop in conductivity, and thus explains the conductivity valley of the $\theta = 3^\circ$ curve at $\epsilon = 88$~meV in Fig.~\ref{fig:cond-intermediate}(b). It is interesting to note that exactly the same band feature is responsible for an \emph{enhancement} of the intraband contribution to the optical conductivity\cite{stau13}, which contrasts to the suppression of in-plane charge conductivity found here. Formally, this results from the different structures of the two theories used to derive these quantities. The optical response is, as with any response function, obtained via second order perturbation theory and the low velocities of the van Hove singularity imply small energy denominators and hence enhancement, whereas in the Boltzmann equation the velocity enters in the numerator and thus its reduction results in a suppression of charge transport.

We next consider the second pronounced feature in $\sigma(\ep)$, the apparently discontinuous drop at $\ep \approx 194\,$meV. As shown in the lower panels of Fig.~\ref{fig:fermisf2.pdf} where we plot the Fermi surfaces either side of this transition ($\epsilon_3 = 193\,$meV to $\epsilon_4 = 195\,$meV), the origin of this feature is somewhat different: we see that a new set of disconnected Fermi sheets arise as the transition point is crossed. This topological change in the Fermiology evidently is (a) discontinuous as the sheets appear at a finite momentum away from the central strongly trigonally warped Fermi surface and (b) results in additional scattering processes, which causes the sudden decrease of the conductivity.

\subsection{Conductivity at small angles: $\theta \le 2^\circ$}
\label{res-con-small}

\begin{figure}[tbp]
  \centering
  \includegraphics[width=0.98\linewidth]{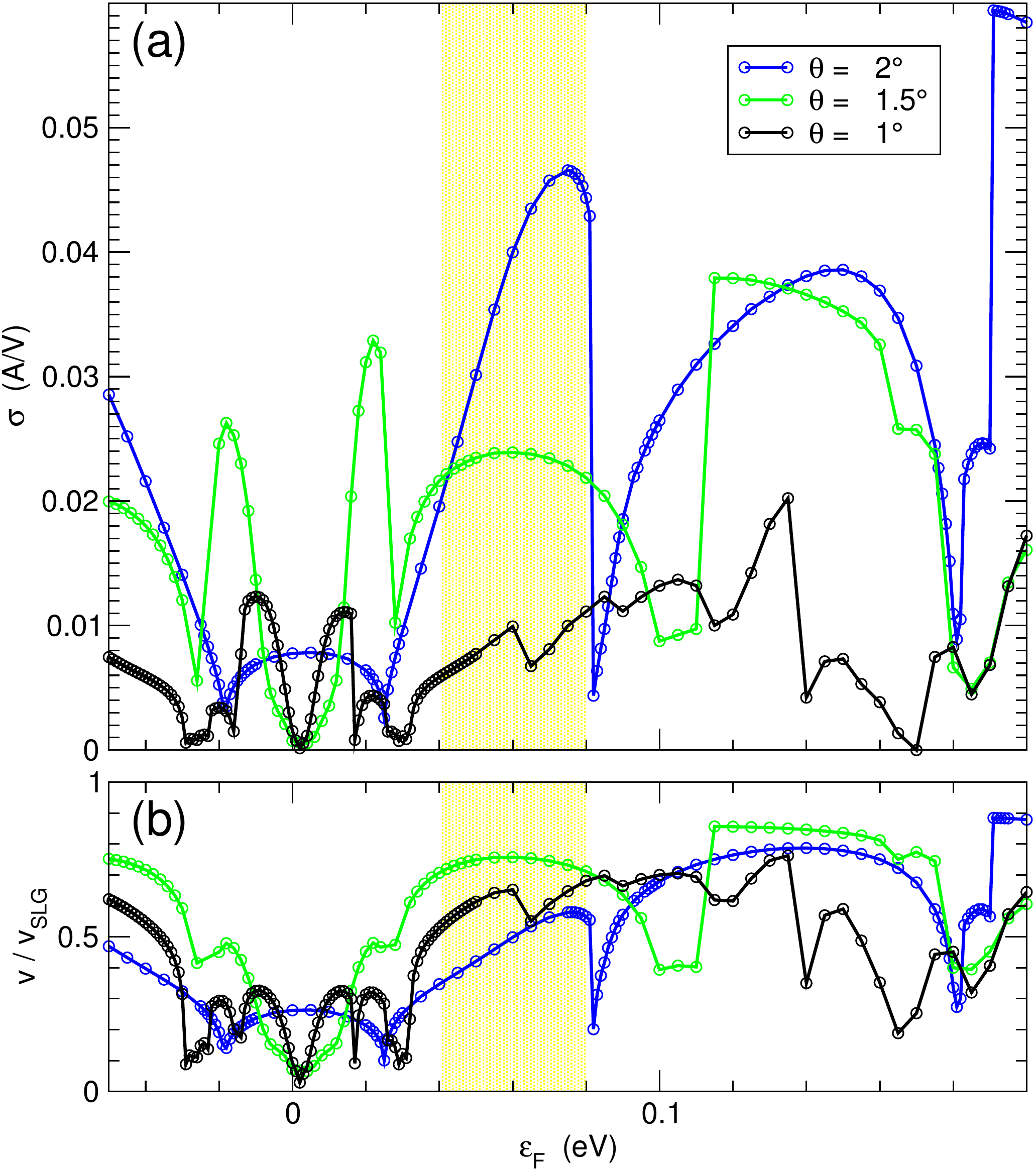}
  \caption{\emph{Importance of the twist bilayer wavefunctions in the small angle regime.} At small twist angles the clear correspondence between the conductivity and the Fermi surface averaged band velocity seen at larger angles no longer holds. To see this compare the Fermi surface averaged velocities (panel (b)) with the conductivities (panel (c)) in the highlighted region. While the average band velocity of the $\theta = 1^\circ$ bilayer in this region is somewhat higher than that for the $\theta = 2^\circ$ bilayer, the conductivity of the former is $\approx$5 times lower than the latter. This indicates that the bilayer wavefunctions now play a dominating role in determining the transport, consistent with the fact that in the small angle limit the twist bilayer wavefunctions become qualitatively different from those of single layer graphene and show features of charge localization\cite{shall13,shall16}. Note that the temperature dependent Fermi smearing is switched off by setting $\partial f_{\vect p}^0/\partial \epsilon_{\vect p}= -\delta(\epsilon_F - \epsilon_{\vect p})$ in Eq.~\eqref{eq:jdefphi} of the main text, and thus the conductivity reflects only scattering processes occurring at the Fermi energy.}
  \label{fig:cond-small}
\end{figure}

In Fig.~\ref{fig:cond-small}(a) we plot the conductivity $\sigma$ as a function of Fermi energy $\epsilon_F$ for a set of rather small rotation angles $\theta$ of the twist bilayer ($\theta = 2^\circ$, $1.5^\circ$, and $1^\circ$). It is immediately apparent that the conductivity at these angles is strongly suppressed as compared to the large angle regime and, furthermore, develops a plethora of non-analytic structures as a function of Fermi energy. These arise from the multiple topological changes in the Fermi surface as a function of energy that characterize the Fermiology of the small angle region\cite{shall16}. Interestingly, it is clear from a comparison between the conductivity and the average band velocity, shown in Fig.~\ref{fig:cond-small}(b), that the transport properties of the bilayer cannot be explained solely on the basis of the average band velocity alone. As a particular example, inspection of the $\theta = 1^\circ$ average velocity in the region close to $60\,$meV shows that while the average band velocity is somewhat higher than that for the $\theta = 2^\circ$ bilayer, the conductivity is 5 times lower. Evidently, the twist bilayer wavefunctions - which enter via the electron-phonon scattering matrix elements $W^{\vect p';\vect p}_{\vect q,\eta}$ - play an crucial role in transport at small angles. This is in contradistinction to the situation at large angles, where the conductivity features could be explained on the basis of the band velocity alone, but is consistent with the fact that, in the small angle regime, the twist bilayer wavefunctions become highly structured due to the interference of many single layer graphene states that are coupled together by the interlayer interaction\cite{shall13}.

\begin{figure}[tbp]
  \centering
  \includegraphics[width=0.98\linewidth]{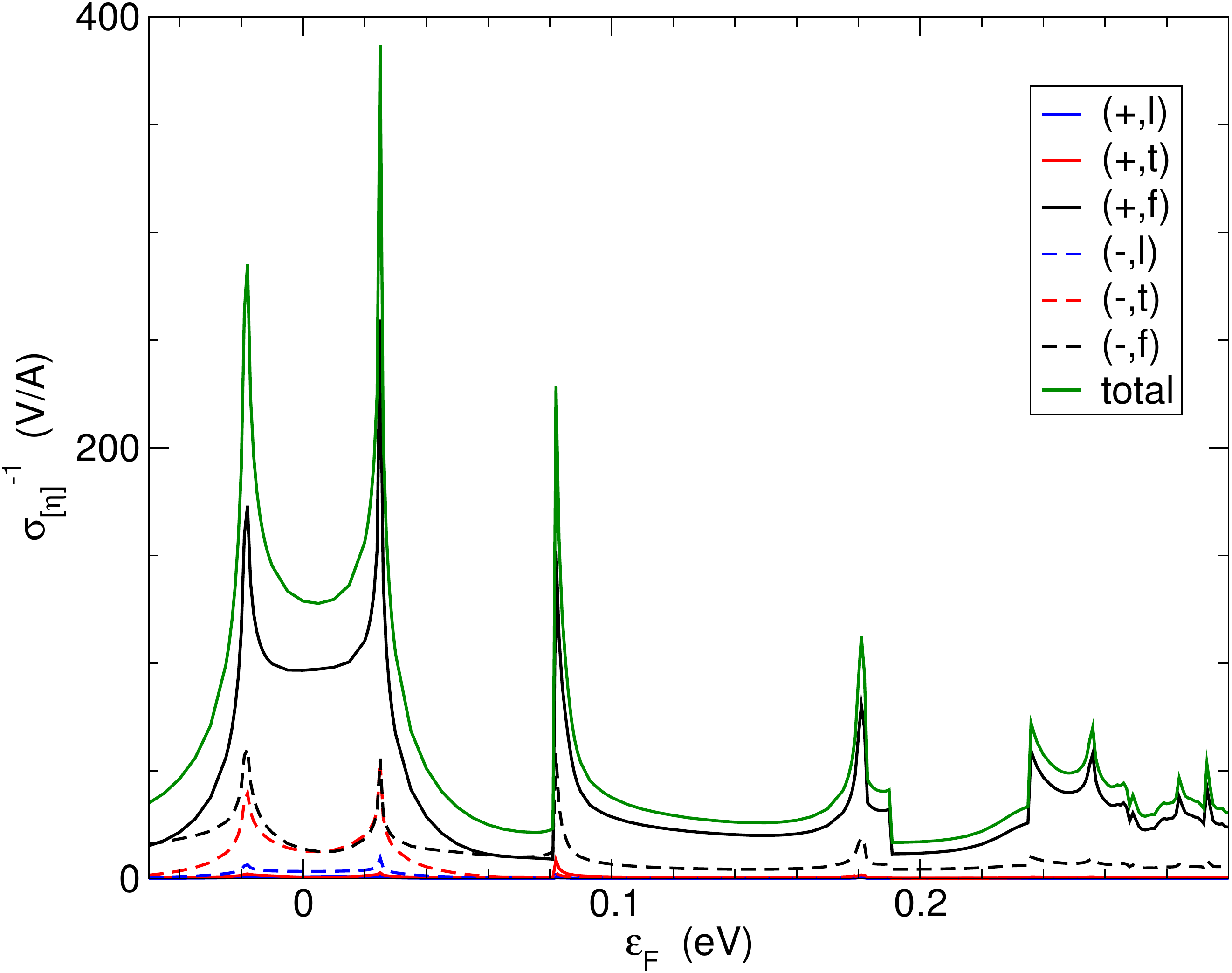}
  \caption{Resistivity $\sigma_{[\eta]}^{-1}$ due to particular phonon modes $\eta=(\sigma,\nu)$ plotted vs. Fermi energy $\epsilon_F$ for temperature $T = 50$~K and rotation angle $\theta=2^{\circ}$. The temperature dependent Fermi smearing $\partial f_{\vect p}^0/\partial \epsilon_{\vect p}$ of the electron states is not included.
}
  \label{fig:phmodes-small.pdf}
\end{figure}

Finally, we investigate the influence of each of the six phonon modes on the small angle conductivity.
In Fig.~\ref{fig:phmodes-small.pdf} we plot the inverse conductivities, i.e. resistivities, $(\sigma_{[\eta]})^{-1}$ for a rotation angle $\theta = 2^{\circ}$ and a temperature $T = 50$~K. We find that each resistivity contribution qualitatively follows the features exhibited by
the total resistivity, but that the resistivity is dominated by the layer-symmetric flexural mode $(+,\text{f})$. At energies $\epsilon_F > 80$~meV it accounts for around 75\% of the total resistivity, while the remaining scattering results almost entirely from the layer-antisymmetric flexural phonon mode $(-,\text{f})$ and each of the in-plane phonon modes contributes only around 1\% to the resistivity. This is in interesting contrast to the large angle regime in which the conductivity was dominated by the layer anti-symmetric flexural mode.


\subsection{Temperature dependence of the conductivity}

\begin{figure}[tbp]
  \centering
  \includegraphics[width=0.98\linewidth]{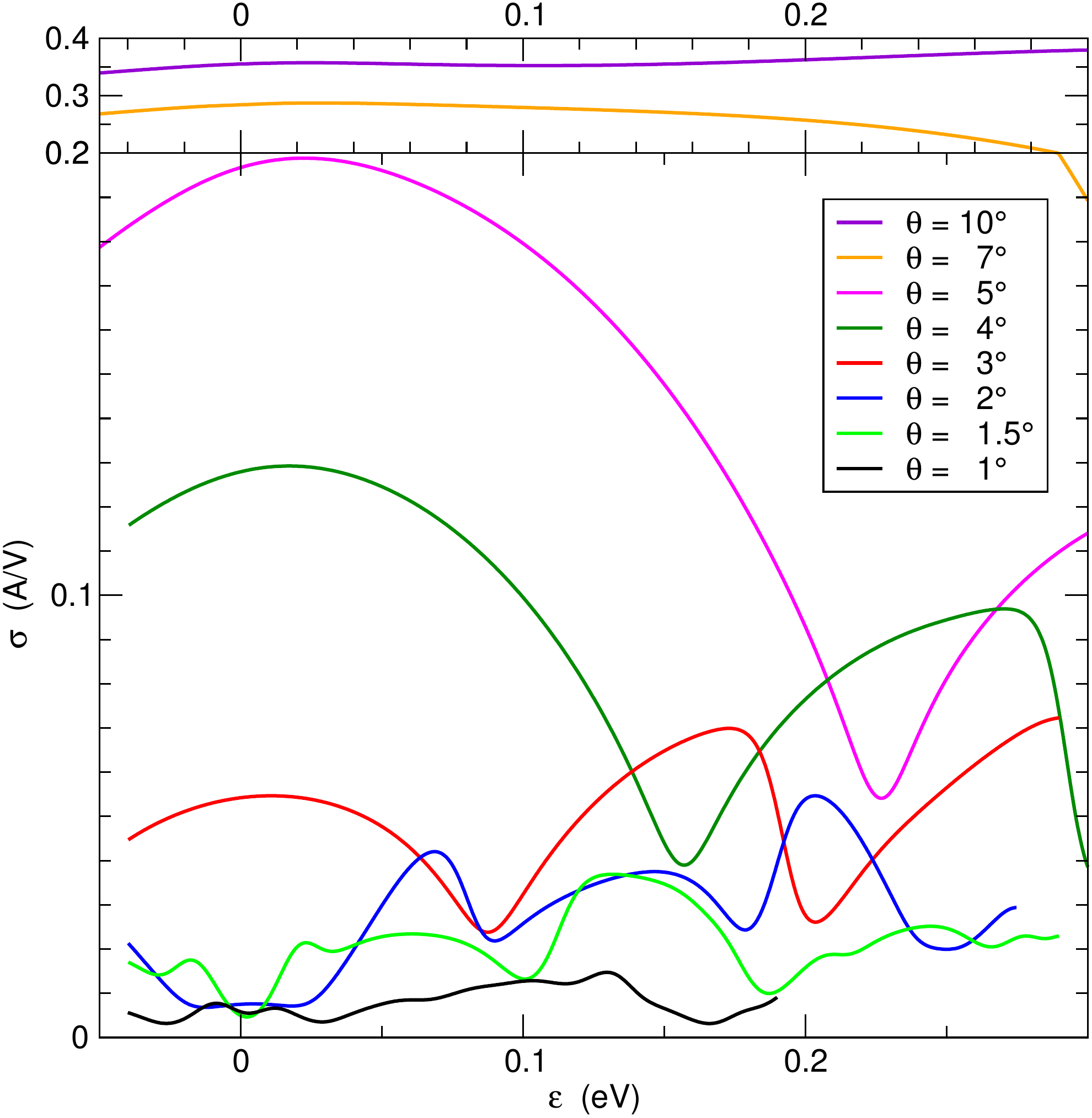}
  \caption{Conductivity $\sigma$ of the twist bilayer as a function of Fermi energy $\epsilon_F$ for a temperature $T = 50$~K and a series rotation angles $\theta$ that encompass both large and small angle cases. The temperature dependent Fermi smearing $\partial f_{\vect p}^0/\partial \epsilon_{\vect p}$ of the electron states is included, which results in a smearing out of the data presented in Fig.~\ref{fig:cond-intermediate}(b) and Fig.~\ref{fig:cond-small}(a), in which the Fermi smearing is not included. Note that the $y$-axis scaling changes at $\sigma = 0.2$~A/V.}
  \label{fig:fdcond.pdf}
\end{figure}

\begin{figure}[tbp]
  \centering
  \includegraphics[width=0.98\linewidth]{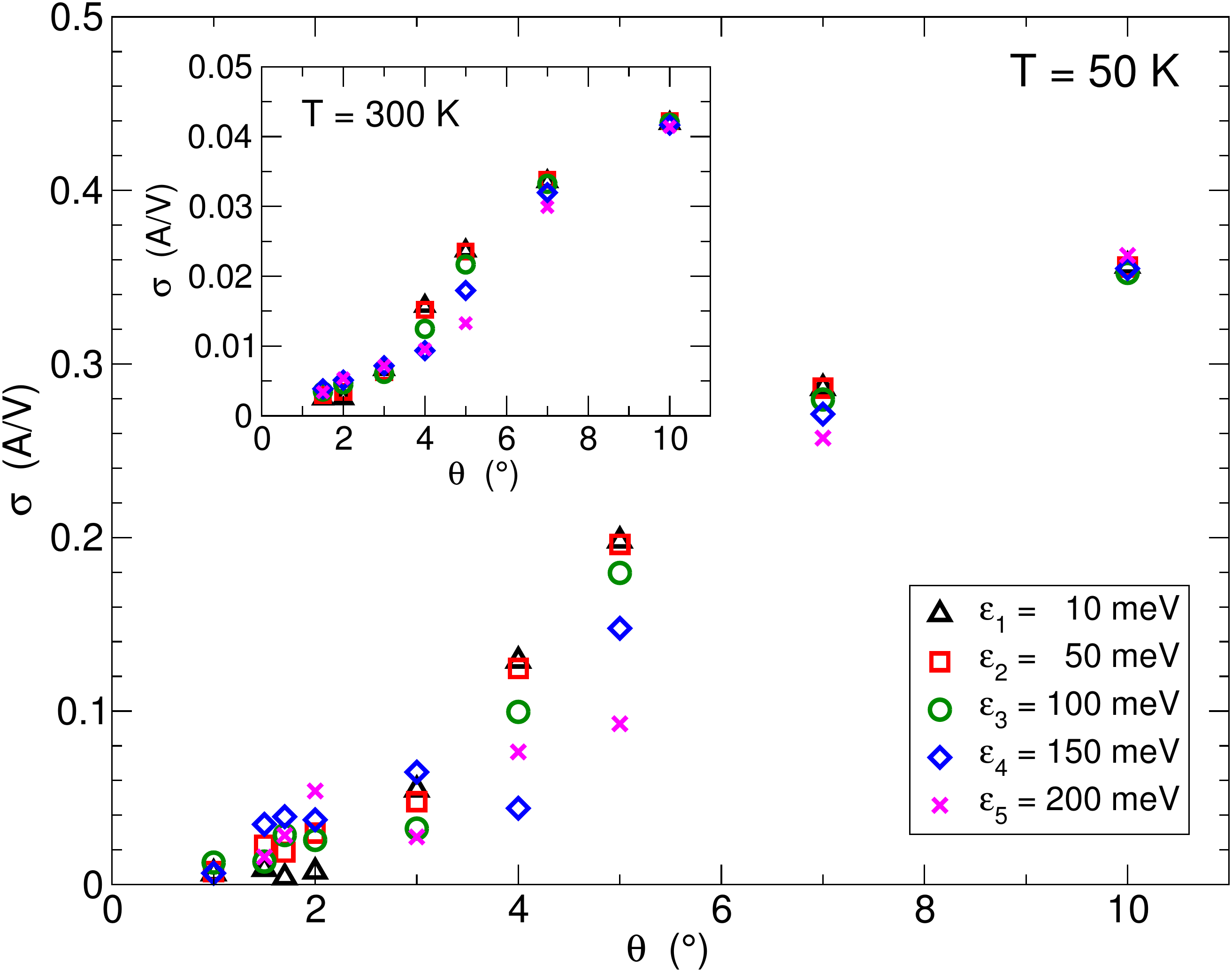}
  \caption{The reduction of conductivity $\sigma$ with decreasing rotation angle $\theta$ of the twist bilayer, shown for a series of Fermi energies $\epsilon_1 = 10$~meV, $\epsilon_2 = 50$~meV, $\epsilon_3 =100$~meV, $\epsilon_4 = 150$~meV and $\epsilon_5 = 200$~meV at temperature $T = 50$~K.
The inset shows the same data but calculated at $T=300$~K.
}
  \label{fig:cvsth.pdf}
\end{figure}

Having described the transport properties of the twist bilayer, and elucidated the relation between the conductivity and the underlying electronic structure, we now restore the Fermi smearing temperature dependence $\partial f_{\vect p}^0/\partial \epsilon_{\vect p}$ in Eq.~\eqref{eq:jdefphi}. In Fig.~\ref{fig:fdcond.pdf} we plot the conductivities for all rotation angles studied in sections \ref{res-con-large}-\ref{res-con-small} as a function of Fermi energy with the temperature set to $T = 50$~K. In comparison to Fig.~\ref{fig:cond-intermediate}(b) and Fig.~\ref{fig:cond-small}(a) we observe that the sharp valleys and peaks are to some extent smoothed due to the Fermi smearing of the order of $k_BT = 4.3\,$meV. The general features, however, are retained (and the intermediate angle non-analytic behaviour of $\sigma(\epsilon_F)$ can be seen at all temperatures $T < 300$~K). In particular, the pronounced reduction in conductivity as a function of twist angle is not, as expected, substantially impacted by restoring the Fermi temperature. In Fig.~\ref{fig:cvsth.pdf} we plot the conductivity
for a series of energies $10\,$meV$\ < \ep_F < 200\,$meV as a function of twist angle for $T = 50\,$K, while the inset shows the same data for $T = 300\,$K.
The general trend of resistivity reduction with twist angle can be seen, but also interestingly it appears that the intermediate angle regime is associated with a large scatter with respect to energy, not found in either the large or small angle regimes.

\begin{figure}[tbp]
  \centering
  \includegraphics[width=0.98\linewidth]{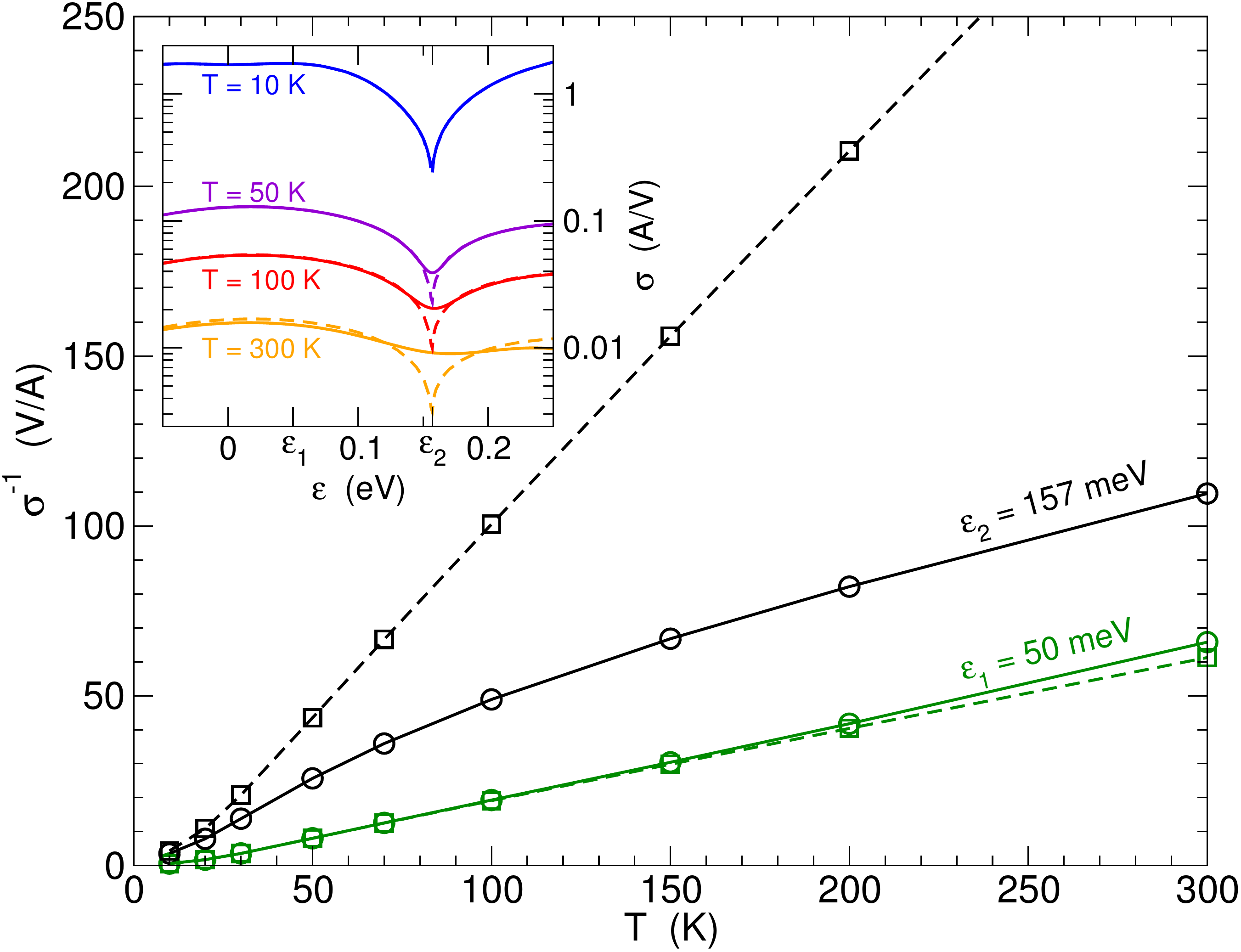}
  \caption{Temperature dependence of the conductivity $\sigma$ of the twist bilayer for a rotation angle of $\theta = 4^\circ$. The main plot shows the inverse conductivity (resistivity) at fixed Fermi energies $\epsilon_1 = 50$~meV and $\epsilon_2 = 157$~meV. The lines are presented only as a guide to the eye. The inset panel shows the conductivity vs. Fermi energy for different temperatures, note the logarithmic scaling of the $y$-axis. In both plots circles (squares) and continuous (dashed) lines represent the results of a calculation including (not including) the temperature dependent Fermi smearing $\partial f_{\vect p}^0 / \partial \epsilon_{\vect p}$.}
  \label{fig:condtmp.pdf}
\end{figure}

Finally we examine the temperature dependence of the conductivity. For a rotation angle of $\theta = 4^\circ$ we consider two representative energies $\epsilon_1 = 50$~meV situated in a region where $\sigma(\ep)$ is approximately a constant function, and $\epsilon_2 = 157$~meV that is situated at the node of the conductivity (see the inset in Fig.~\ref{fig:condtmp.pdf}). In the main plot of Fig.~\ref{fig:condtmp.pdf} we present the temperature dependence of the inverse conductivity $\sigma^{-1}$ for these two energies. In both plots are shown (i) the results of a calculation without the temperature dependent Fermi smearing of the electron distribution (dashed lines and squares), i.e. $\partial f_{\vect p}/\partial \epsilon_{\vect p} := -\delta(\epsilon_F-\epsilon_{\vect p})$, as well as (ii) the results of a calculation including the Fermi smearing (continuous lines and circles). With the only temperature dependence arising from the phononic contribution an increase in temperature will result in an overall lower conductivity due to the increased phonon population. With only one phonon mode with fixed energy $\omega$, one would find $\sigma \sim \beta n_{\omega} (1+n_{\omega})$, which for $k_B T \gg \omega$ yields $\sigma \sim T^{-1}$. The dashed lines in Fig.~\ref{fig:condtmp.pdf} show this linear temperature dependence of the inverse conductivity for $T > 30$~K. Including additionally the Fermi smearing of the electron distribution (continuous lines in Fig.~\ref{fig:condtmp.pdf}) allows electrons of a wider range of energies to participate in transport and hence is particularly relevant in an energy region where the conductivity contribution from constant energy surfaces changes rapidly with energy, e.g. at $\epsilon_2$ in Fig.~\ref{fig:condtmp.pdf}. As may be seen in the main plot the $\epsilon_2$ conductivity data is dramatically changed by the inclusion of Fermi smearing. However, a normal temperature dependence (i.e., an increase of the resistivity with temperature is always observed).


\subsection{Conductivity dependence on a layer perpendicular electric field}
\label{res-efp}

The rich Fermiology of the small angle limit of the twist bilayer arises due to the fact that the momentum scale on which the hybridization of the bare Dirac cones takes place becomes small in this limit (recall the moir\'e momentum is proportional to $\sin\theta/2$). This suggests that a displacement of the bare Dirac cones by, for instance, an applied interlayer bias, will lead to significant changes in the electronic structure of the small angle twist bilayer and,
possibly, 
in the transport properties. In this way we may imagine that transport in the small angle limit may be particularly susceptible to external perturbation. In this section we will investigate this via application of a layer perpendicular electric field $\vect E_{\perp} = E_{\perp} \hat{\vect z}$ that shifts the bare Dirac cones by $\lambda ce E_{\perp}/2$ ($\lambda = \pm$ labels the layers, $c$ is the interlayer distance).

\begin{figure}[tbp]
  \centering
  \includegraphics[width=0.98\linewidth]{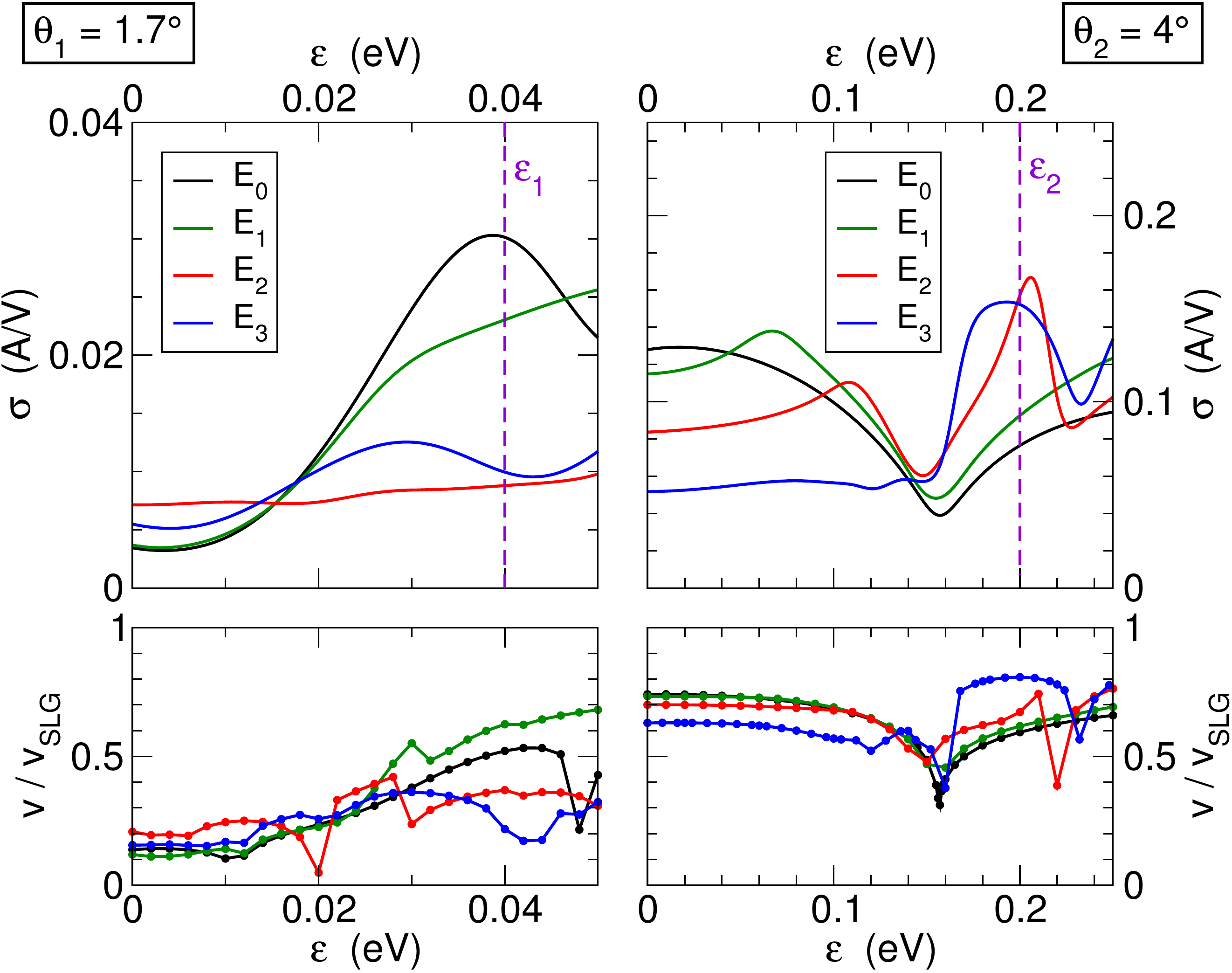}
  \caption{Conductivity $\sigma$ (top) and average band velocity $v$ (bottom) plotted versus Fermi energy $\epsilon_F$ for temperature $T = 50$~K and the two rotation angles $\theta_1=1.7^{\circ}$ (left) and $\theta_2=4^{\circ}$ (right). The different colors correspond to different strengths of the perpendicular electric field: $E_0 = 0$, $E_1 = 59.7 \ \text{mV}/\text{\AA}$, $E_2 = 119.4 \ \text{mV}/\text{\AA}$ and $E_3 = 179.1 \ \text{mV}/\text{\AA}$. The temperature dependent Fermi smearing $\partial f_{\vect p}^0/\partial \epsilon_{\vect p}$ of the electron states is included.}
  \label{fig:perpe.pdf}
\end{figure}

In Fig.~\ref{fig:perpe.pdf} we display the conductivity $\sigma$ and Fermi surface averaged band velocities $v$ for different values of $E_{\perp}$ for the two rotation angles $\theta_1 = 1.7^{\circ}$ and $\theta_2 = 4^{\circ}$. We choose a series of field strengths $E_j = j\cdot 59.7  \ \text{mV}/\text{\AA}$, with
$j = 0,...,3$. 
This correspond to potential differences of $\Delta\Phi_j = j\cdot 0.2$~V between the two layers. As may immediately be seen, at energies $\epsilon_1 = 40$~meV for $\theta_1$ and $\epsilon_2=200$~meV for $\theta_2$, the conductivity changes by up to a factor of $\approx 3$.

\begin{figure}[tbp]
  \centering
  \includegraphics[width=0.98\linewidth]{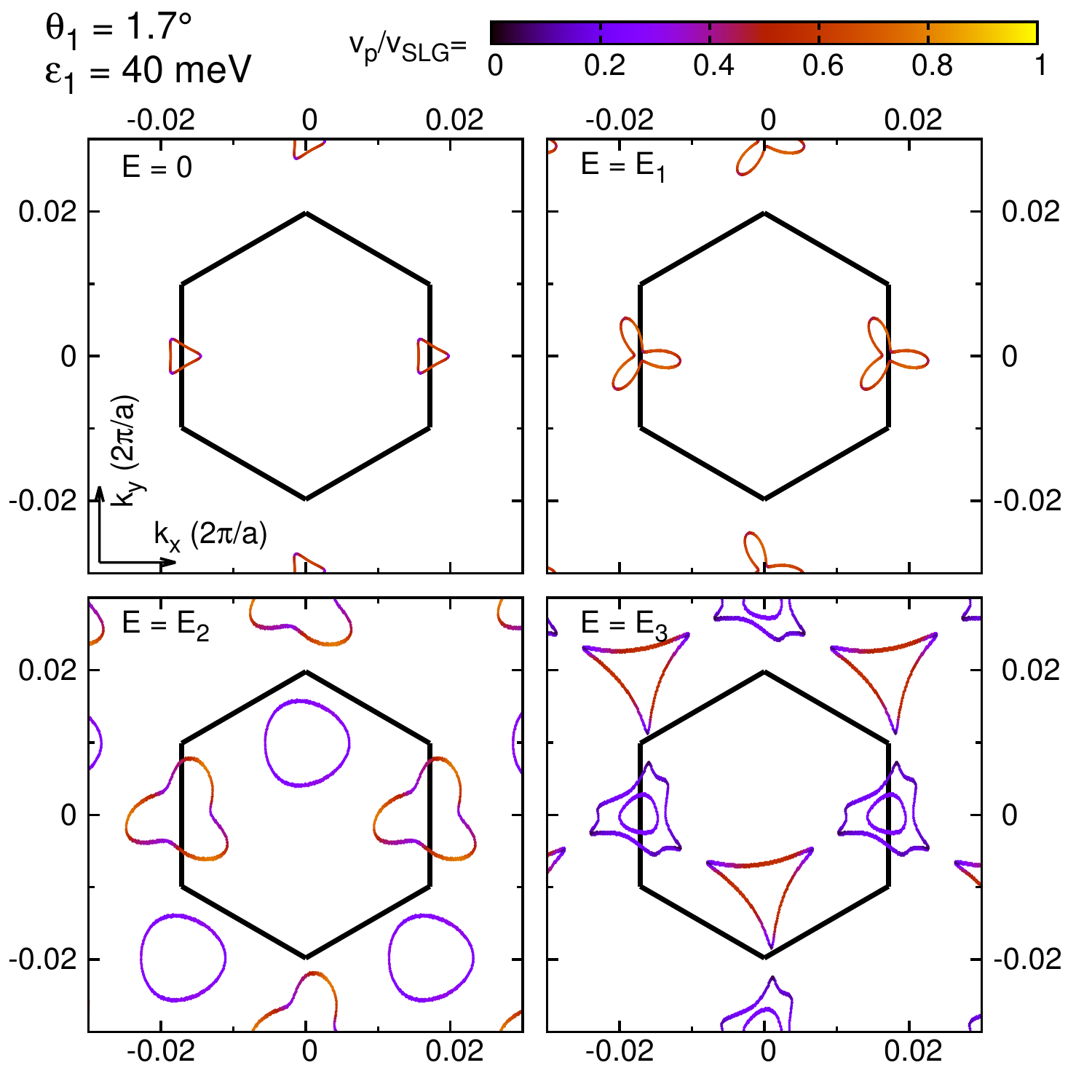}\\
  \includegraphics[width=0.98\linewidth]{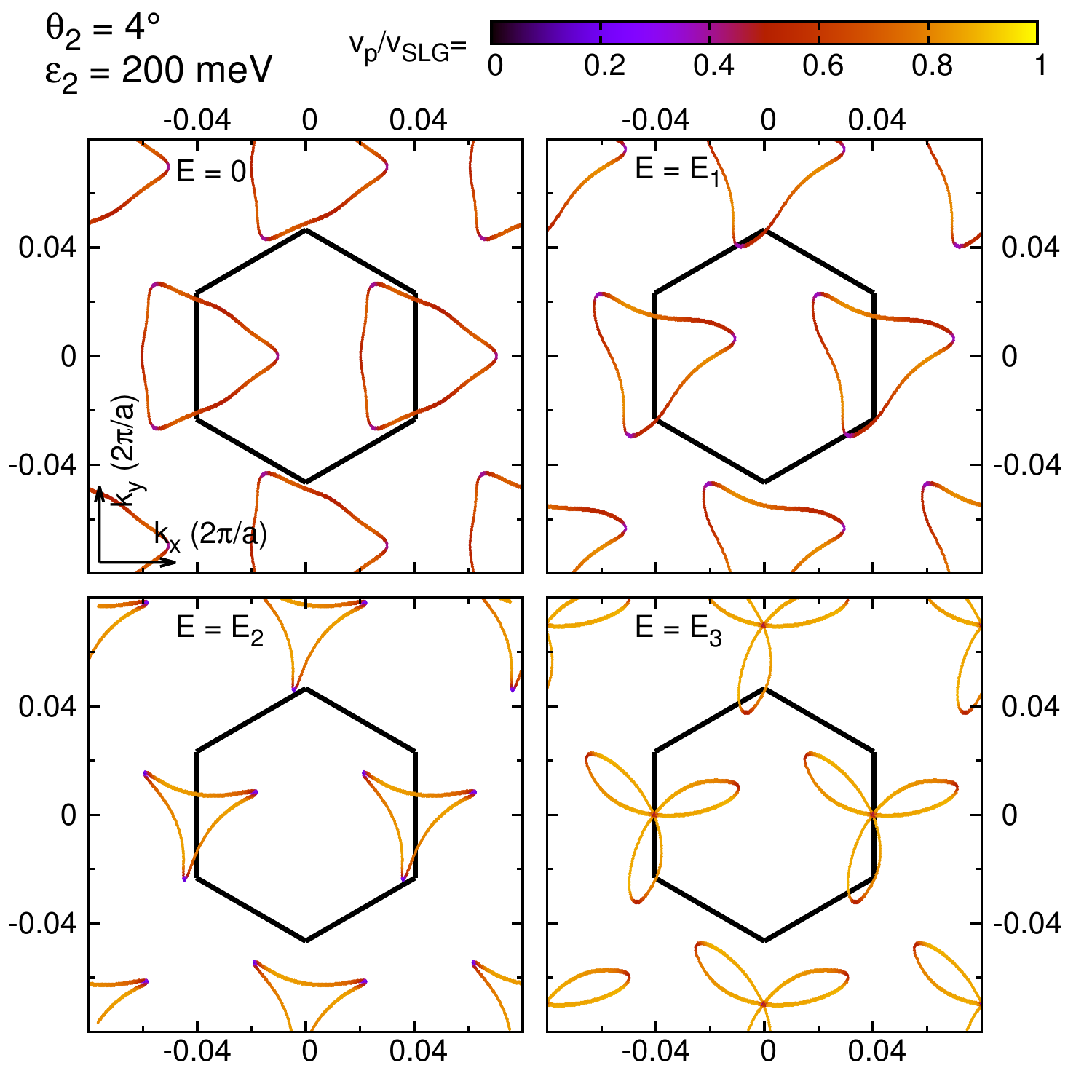}
  \caption{Fermi surfaces at $\theta_1 = 1.7^\circ$, $\epsilon_1 = 40$~meV and at $\theta_2 = 4^\circ$, $\epsilon_2 = 200$~meV for different strength of the perpendicular electric field: $E_0 = 0$, $E_1 = 59.7 \ \text{mV}/\text{\AA}$, $E_2 = 119.4 \ \text{mV}/\text{\AA}$ and $E_3 = 179.1 \ \text{mV}/\text{\AA}$. The color encodes the band velocity $|\vect v_{\vect p}|$ in units of the SLG Dirac band velocity $v_{\text{SLG}}$, the black hexagons depict the twist bilayer reciprocal unit cell.}
  \label{fig:perpefs}
\end{figure}

To explain these changes in the bilayer conductivity we analyze the corresponding Fermi surfaces, which we display in Fig.~\ref{fig:perpefs}. In the case of $\theta_1$ and $\epsilon_1$ with increasing electric field new low velocity Fermi sheets appear allowing for more scattering and therefore decreasing the conductivity. In the case of $\theta_2$ and $\epsilon_2$ the band velocity increases with increasing perpendicular field, which in this case leads to an increase in the conductivity until the field reaches $E_2$. Further increase of the perpendicular field to $E_3$ leads to a further increase in the average band velocity, however the additional significant change in the shape of the Fermi surface alters the allowed scattering processes such that that the conductivity does not change. These results suggest that the bilayer may indeed form an interesting system for manipulation of transport properties by external perturbation, although it should be stressed that the results presented here are smeared out upon increasing the temperature above 50~K.


\section{Conclusions}
\label{conclusion}

We have surveyed the in-plane electric conductivity of the graphene twist bilayer in a wide range of twist angles $1^\circ < \theta < 30^\circ$ and Fermi energies $\epsilon < 300\,$meV. The calculations have been performed on the basis of an effective Hamiltonian for the twist bilayer band structure, first introduced in Ref.~\onlinecite{shall16}. The transport problem has been treated by (i) employing the model of an isotropic elastic bilayer for the phonon dispersion and (ii) using the linear Boltzmann equation for elastic electron-phonon scattering to calculate scattering probabilities and the conductivity.

Similar to the ground state of the twist bilayer we find that the in-plane transport properties are qualitatively different in three distinct angle regimes. At large twist angles $10^\circ < \theta < 30^\circ$ the conductivity may be characterized by an \emph{interlayer Bloch-Gr\"uneisen temperature}: below this temperature phonons of sufficient momentum to scatter between the Dirac cones of the mutually rotated layers do not exist, and the bilayer is decoupled (with the total conductivity simply a sum of the conductivities of the two layers). Above this temperature, even though in the ground state the bilayer is decoupled, the transport problem re-couples the bilayer. In particular the conductivity, in striking contrast to single layer graphene, does not possess particle-hole symmetry. This arises from the fact that the effective moir\'e potential that describes the coupling between the two layers of the twist bilayer (a complex valued $\br$-dependent field) does not possess particle-hole symmetry.

At intermediate angles $3^\circ < \theta < 10^\circ$ two sharp transitions are seen in the energy dependence of the conductivity, which otherwise presents a smooth function. The first of these transitions is related to the well known van Hove singularity that occurs at the energy for which the cones from each layer first intersect, and results in a pronounced drop in the Fermi velocity and hence conductivity. The second sharp transition in the conductivity arises due to a topological change in the Fermi surface -- a Lifshitz transition -- that occurs at the energy at which back-folded bands to the effective moir\'e Brillouin zone create new electron pockets that trigonally decorate the strongly warped Dirac cone. The increased scattering to these low velocity sheets causes a rapid reduction of the conductivity.

At very small angles of $\theta < 2^\circ$ the conductivity is suppressed by almost two orders of magnitude compared to the large angle case and, furthermore, develops a richly structured energy dependence. In contrast to the large and intermediate angle conductivity, where features of the conductivity can be clearly related to corresponding features of the Fermi surface averaged band velocity, we find this is not the case in the small angle regime. The reason for this difference is that at large and intermediate angles the twist bilayer wavefunctions are very close to those of single layer graphene, and therefore the changes in conductivity are dominated by band structure changes that the interlayer interaction induces. However, the significantly stronger interlayer interaction in the small angle regime results both in renormalization of the band structure, as well as in wavefunctions that differ qualitatively from those of single layer graphene, and indeed show pronounced features of charge localization\cite{shall13,shall16}. The scattering of such twist bilayer states by phonons will evidently be very different to the phonon scattering of states that can be well approximated as single layer graphene states, and it is this effect that is responsible for the loss of a clear correspondence between the Fermi surface averaged band velocity and the conductivity.

Finally, we have addressed the issue of how the conductivity of the bilayer may be manipulated by an external layer perpendicular field, finding that it is possible to do so, but that at room temperature such effects will be largely washed out by Fermi smearing. Future work on this interesting system should address the role that impurities play in both the electronic structure and transport of the twist bilayer. In particular, in the small angle regime in which the bilayer wavefunctions differ qualitatively from those of single layer graphene, the physics of impurity scattering may be very different from that of either graphene or Bernal stacked bilayer graphene.

\begin{acknowledgments}

This work was supported by the Collaborative Research Center SFB 953 of the Deutsche Forschungsgemeinschaft (DFG) and by the DFG grant PA 516/8-1.

\end{acknowledgments}


%

\end{document}